\documentclass[aps,prb,twocolumn,showpacs,superscriptaddress]{revtex4}
\usepackage{graphicx}
\usepackage{epstopdf}
\usepackage{latexsym}
\usepackage{amssymb}
\usepackage{amsmath}
\usepackage{amsfonts}
\usepackage{subfigure}
\usepackage{bm}
\usepackage{multirow}

\newcommand{\Tr}{\mathop{\mathrm{Tr}}}
\newcommand{\sgn}{\mathop{\mathrm{sgn}}}
\newcommand{\vect}[1]{{\bm{#1}}}
\renewcommand{\Im}{\mathop{\mathrm{Im}}}
\newcommand{\eqnref}[1]{Eq.\,\eqref{#1}}
\newcommand{\figref}[1]{Fig.\,\ref{#1}}
\newcommand{\tabref}[1]{Tab.\,\ref{#1}}
\newcommand{\diag}[2]{\vcenter{\hbox{\includegraphics[height=#2]{#1}}}}

\begin{document}

\title{Two-Fluid Description for Iron-Based Superconductors}
\author{Yi-Zhuang You}
\affiliation{Institute for Advanced Study, Tsinghua University, Beijing, 100084, China}
\author{Zheng-Yu Weng}
\affiliation{Institute for Advanced Study, Tsinghua University, Beijing, 100084, China}
\date{\today }

\begin{abstract}
We present a two-fluid description for iron-based superconductors, which contains an itinerant electron Fermi-liquid and a local moment spin-liquid, coupled together via an effective Hund's rule interaction. We examine the low-energy collective behavior of such a system. We find that an electron-spinon composite mode emerges in the intermediate coupling regime, which may account for the hump-dip behavior observed in the recent scanning tunneling spectroscopy experiments. The superconductivity and spin-density-wave phases are consistently described within the same framework. Possible  experimental test is also proposed.
\end{abstract}


\maketitle

\section{Introduction}

The discovery of iron-based superconductors\cite{Hosono08} has triggered a new round of intensive research of high-temperature superconductivity. One early dispute focused on the attribute of the iron $3d$ electrons: whether they are itinerant quasi-particles like those in metals, or localized magnetic moments like those in Mott insulators. Underlying this dichotomy between the itinerant electron and the local moment are the two different schools of thoughts for superconductivity (SC): the weakly correlated Bardeen-Cooper-Schrieffer\cite{BCS} (BCS) theory based on the electron Fermi-liquid, and the strongly correlated resonant valence bound\cite{RVB} (RVB) theory arising from the doped Mott insulator.

After five years of experimental and theoretical study, today it seems that the consensus is converging\cite{Mazin09,Zhao09,Hansmann10,Gretarsson11, Vilmercati12, Gorkov13} to the idea of coexistence of both itinerant and local degrees of freedom. Due to the multi-orbtial (multi-band) nature, the orbital-selective Mott (OSM) transition\cite{Medici09, Lv10, Medici11, Yu11,Zhang:2012bh, Quan:2012dq, Yu:2012ve} may take place in the iron-based compounds, allowing the electrons to play multiple roles at the same time. Based on the experimental evidences,\cite{Moon10,NLWang12}  several versions of the itinerant electrons and local moments hybrid theory were proposed by the authors of Ref.\,\onlinecite{Weng09, KouLiWeng09,Yin10}, and its rich phase diagram was studied in Ref.\,\onlinecite{YouPRB11}. In this scenario, the Fe $d$-orbital electrons are separated into two well-defined degrees of freedom: itinerant electrons and local moments, and they are coupled together through the residual on-site Hund's rule interaction. The iron-based SC is understood as the pairing of itinerant electrons glued by the (para-)magnon of local moments, while the spin-density-wave (SDW) order in the parent compounds is considered as a joint ordering of both degrees of freedom whose ordering tendencies are mutually enhanced by the Hund's rule coupling.

Out of the SDW phase, the local moments were assumed to be in a disordered paramagnetic ground state, which is compatible with the non-magnetic SC phase at low temperature. However, motivated by the recent experimental discovery of the gap-like hump-dip feature in the normal phase,\cite{YYWang12,Wang2013} we conjecture that the local moments may be effectively described by a bosonic spin-liquid state, with gapped deconfined spinons, which is important for a consistent explanation (to be discussed in the following) of the observed hump-dip feature.

Based on this assumption, we develope a two-fluid description for iron-based superconductors, which contains two liquid components: the itinerant electron \emph{Fermi-liquid} and the local moment \emph{spin-liquid}. The spin-liquid\cite{Wen91, ReadSachdev91} physics is characterized by a mean-field theory for spinons,\cite{Sachdev94, Senthil04} which are the low-energy excitations of the spin-liquid. Here we assume bosonic spinons,\cite{Arovas88} which are then coupled to the fermionic itinerant electron via the residual Hund's rule interaction, making the system a Bose-Fermi mixture. Within the minimal model, we examine the Fermi-liquid instability in coupling with the spin-liquid, the resonance modes between the two fluids, and the low-energy collective modes originated from the coupling. The main discovery was that in the intermediate coupling region,  there exists a composite fermion bound-state made up of an electron and a spinon, which is close to but gapped from the Fermi surface and carries unit electron charge and integer spin. We proposed that such a composite mode may account for the hump-dip feature\cite{YYWang12, Wang2013} in the scanning tunneling spectroscopy (STS) experiments.

Admittedly the two-fluid description is just a low-energy effective theory about the interplay between the spin and the charge degrees of freedoms in the iron-based compounds, but not intended to provide a complete or realistic modeling of the material. For example, we have omitted the orbital degrees of freedom, which has been found to be another important ingredient of the material by many experiments.\cite{Davis10, Fisher10, Shen11, Xue11, Matsuda12, Fisher12} As we mainly focus on the SDW and SC phases in the phase diagram, where the orbital fluctuation plays a minor effect,\cite{FYang2013} the two-fluid description should be sufficient to capture the main physics in many aspects. The incorporation of the orbital physics will be left for future research. 

The remaining of the paper is organized as follows. In section II, we will introduce the two fluid description in the Hamiltonian formalism, by setting up the models for the Fermi-liquid and the spin-liquid respectively. The model is then summarized and reformulated in the path integral formalism in section III. The Hund's rule interaction couples the two fluid components together to produce low-energy collective excitations: magnons and composite fermions, which are discussed respectively in section IV and V. Concluding remarks are given in section VI.

\section{Two-Fluid Discription}
\subsection{Coexistence of Itinerant Electron and Local Moment}
In this section, we will briefly review the itinerant electron and local moment hybrid model, based on which, the two-fluid description will be introduced in the following sections. The hybrid model\cite{KouLiWeng09,YouPRB11} is given by the Hamiltonian
\begin{equation}
H = H_\text{itn}+H_\text{loc} +H_\text{int}.
\end{equation}
$H_\text{itn}$ describes the itinerant electrons near the Fermi surface hopping on the Fe square lattice
\begin{equation}\label{eq: H_c TBM}
H_\text{itn} = \sum_{i,j;\alpha,\beta;\sigma}t_{ij}^{\alpha\beta}c_{i\alpha\sigma}^\dagger c_{j\beta\sigma} + h.c.,
\end{equation}
where $c_{i\alpha\sigma}$ denote the fermion operators for itinerant electrons, with $i,j$ labeling the Fe sites, $\alpha,\beta$ labeling the itinerant orbitals (bands) and $\sigma = \uparrow,\downarrow$ labeling the electron spins. $t_{ij}^{\alpha\beta}$ is some tight-binding model parameters that can be determined from the local density approximation (LDA) calculation or by fitting to the angular-resolved photoemission spectroscopy (ARPES) observations. $H_\text{loc}$ describes the exchange coupling between local moments in the orbital-selective Mott bands
\begin{equation}\label{eq: H_b MM}
H_\text{loc} = \sum_{i,j}J_{ij}\vect{M}_i\cdot \vect{M}_j,
\end{equation}
where $\vect{M}_i$ stands for the local moment on each site. The moment size is roughly spin-1, as suggested by the X-ray emission spectroscopy (XES) observations.\cite{Gretarsson11, Vilmercati12} . Here the local moment, though maybe large, is an effective one presumably dressed by the higher energy itinerant  electrons, which is not quite well quantized, and may be still subject to much larger quantum fluctuations than the usual Heisenberg spins. $J_{ij}$ describes the effective exchange coupling strength between site $i$ and site $j$, which could be determined by fitting its resulting spin wave spectrum to the inelastic neutron scattering (INS) observations. Many have proposed the $J_1$-$J_2$ model\cite{Si08, MaLuXiang08, Sigh10, Si11} or its variants as the magnetic model for the iron-based compounds, which contains nearest-neighbor $J_1$ and next-nearest-neighbor $J_2$ exchange couplings. For large $J_2/J_1$, the $J_1$-$J_2$ model can give rise to the collinear magnetic ordering \figref{fig: J1J2}(a), as observed in many  compounds.\cite{Cruz08, Huang08, Li09} However around $J_2/J_1\simeq0.5$, the model can be strongly frustrated, and may support the spin-liquid state (a paramagnetic quantum ground state).\cite{Richter08, Jiang09,Richter10, Reuther10, Jiang11, LWang11,TLi12,YangYao12} Moreover, the competing Ruderman-Kittel-Kasuya-Yosida (RKKY) (or double-exchange) interaction introduced by the itinerant electron\cite{Yin10,Yin12} and the ring exchange interaction in the intermediate correlated system\cite{Yao10} both tend to stabilize the spin-liquid state. $H_\text{int}$ describes the residual ferromagnetic Hund's rule interaction\cite{Lv10} between itinerant electrons and local moments
\begin{equation}\label{eq: H_int SM}
H_\text{int} = -\sum_{i,\alpha}J_{H} \vect{S}_{i\alpha} \cdot \vect{M}_i,
\end{equation}
where $\vect{S}_{i\alpha}=\frac{1}{2}c_{i\alpha\sigma}^\dagger \vect{\sigma}_{\sigma\sigma'}c_{i\alpha\sigma'}$ is the spin operator for itinerant electrons, with $\vect{\sigma}=(\sigma_1, \sigma_2, \sigma_3)$ being the three Pauli matrices, not to be confused with the spin index. $J_{H}>0$ represents the Hund's rule coupling strength.

At a first glance, the hybrid model looks similar to the Kondo lattice model\cite{Doniach79, Lacroix79} for the heavy-fermion system, both having the itinerant electrons magnetically coupling to the local moments. In fact, D. Pines and his collaborators\cite{Nakatsuji04, Yang08hc, Yang08bs} have proposed a similar two-fluid description for the Kondo lattice, which states that the system may be viewed as a combination of the heavy fermion fluid and the screened Kondo centers. Our work is inspired by their proposal. However, an important difference between the two models is that the Hund's rule coupling is ferromagnetic, while the Kondo coupling is antiferromagnetic. Under renormalization flow,\cite{Anderson70} the ferromagnetic coupling will become weaker, as opposite to the antiferromagnetic coupling. So there is no Kondo screening in the iron-based compounds. In the low-energy limit, the itinerant electrons and the local moments are well-defined degrees of freedom, and the Hund's rule coupling can be treated as a perturbation. 

Based on the hybrid model, we will establish the minimal effective model for the itinerant electrons and the local moments in the following sections. We propose that the itinerant electrons can be described by the Fermi-liquid, and the local moments can be described by the spin-liquid. The interplay between the two fluid components would give rise to the SDW, SC phases and the hump-dip feature in the normal phase.

\subsection{Fermi-Liquid Description of Itinerant Electrons}
\begin{figure}[htbp]
\begin{center}
\includegraphics[width=0.36\textheight]{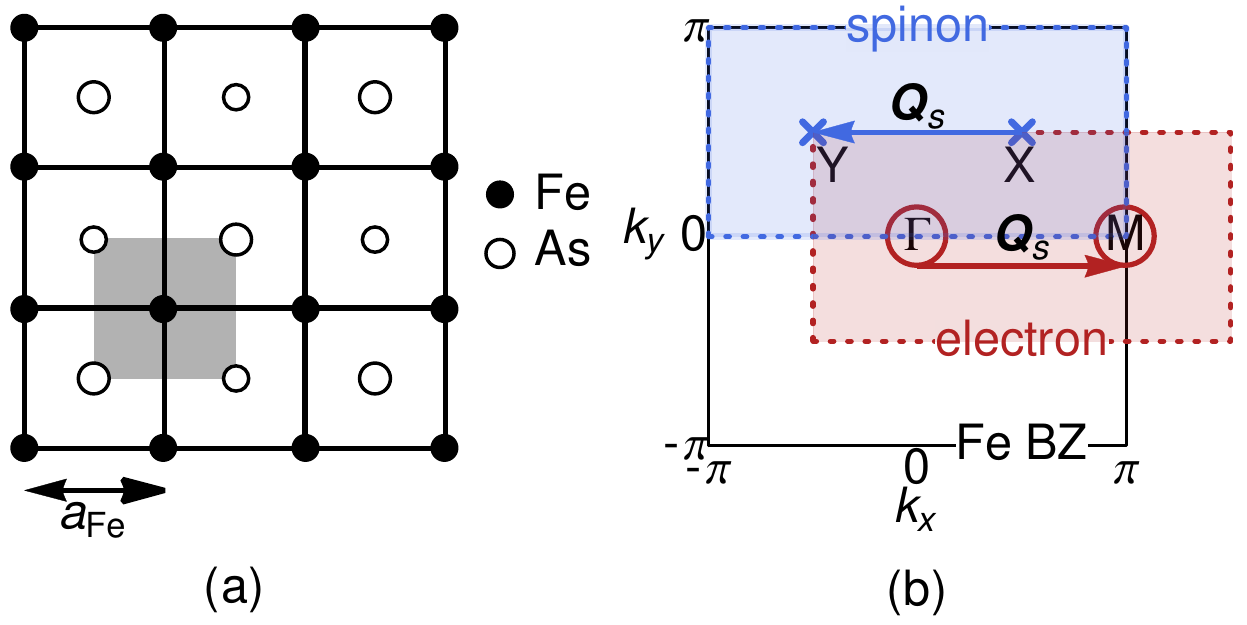}
\caption{(a) Top view of the FeAs layer. The nearest Fe-Fe distance $a_\text{Fe}\simeq 2.8\text{\AA}$ is set to unity through out this article. The shaded area is the Fe unit cell. (b)The Brillouin zones (BZs). The unfolded Fe BZ of Fe lattice is bounded by the solid lines. The folded BZ for the itinerant electron (local moment spinon) is reshaped into the red (blue) rectangle. The momentum is measured in unit of $a_\text{Fe}^{-1}$: $\Gamma(0,0)$, $M(\pi,0)$, $X(\pi/2,\pi/2)$, $Y(-\pi/2,\pi/2)$. The nesting momentum $\vect{Q}_s=(\pi,0)$ connecting the electron pockets (red circles around $\Gamma$ and $M$) is also the ordering momentum connecting the two spinon band softening points (blue crossings at $X$, $Y$).}
\label{fig: lattice BZ}
\end{center}
\end{figure}

Although the iron-based superconductor involves a rather complicated multi-band electronic structure, only the bands near the Fermi level are responsible for electron itineracy. In most families of the iron-based superconductor compounds, it is observed that the Fermi surface usually consists of totally four pockets: two hole pockets around $\Gamma(0,0)$ point and two electron pockets around $M(\pi,0)$ (or equivalently $(0,\pi)$) point\cite{Mazin10,Hirschfeld11} (through out this paper, we will stick to the convention of Fe unit cell and the corresponding Brillouin zone, see \figref{fig: lattice BZ}). The scattering between $\Gamma$ and $M$ pockets turn out to be the most important scattering channel, because the momentum connecting $\Gamma$ and $M$ pockets, i.e.  $\vect{Q}_s=(\pi,0)$, matches the collinear ordering momentum of the local moment, so that the itinerant electron is most strongly scattered with the local moment in this channel.

\begin{figure}[htbp]
\begin{center}
\includegraphics[width=0.2\textheight]{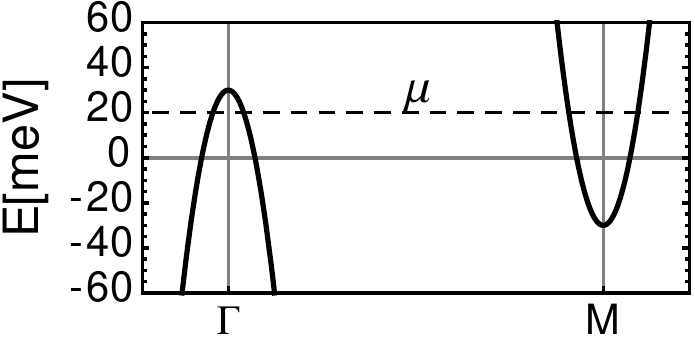}
\caption{Two-pocket band structure of the itinerant electron, with $m_c=0.83$eV$^{-1}$ and $\epsilon_0= -30$meV. Dashed line indicates the Fermi level, determined by the chemical potential $\mu$. The zero-energy level is set to the maximally nested level.}
\label{fig: c band}
\end{center}
\end{figure}

To simplify the problem, we consider the minimal two-pocket model\cite{Chubukov08,Hu12} by keeping only two itinerant bands: one hole band around $\Gamma$ point and one electron band around $M$ point. This model retains the most essential scattering channel between $\Gamma$ and $M$ pockets, making it capable to describe generic phases in iron-based compounds, such as the SDW and the $s_\pm$-wave SC. The extension to the more realistic four-pocket (or even five-orbial) model is straight forward. The Hamiltonian for the two-pocket model is
\begin{equation}\label{eq: H_cc}
H_{c} = \sum_{\vect{K},\vect{k},\sigma}
c_{(\vect{K}+\vect{k})\sigma}^\dagger (\epsilon_\vect{K}(\vect{k})-\mu) c_{(\vect{K}+\vect{k})\sigma},
\end{equation}
which can be considered as the momentum space representation of the tight-binding Hamiltonian $H_\text{itn}$ in \eqnref{eq: H_c TBM}. $c_{\vect{K}+\vect{k}\sigma}$ ($\vect{K}=\Gamma,M$) denotes the operator for itinerant electrons around the $\Gamma,M$ point, $\epsilon_\vect{K}(\vect{k})$ describes the corresponding band dispersion. $\vect{k}$ is the momentum deviation measured from the $\Gamma$ or $M$ point. We may simply take the following parabolic band structure (see \figref{fig: c band}),
\begin{equation}
\epsilon_{\Gamma}(\vect{k})=-\frac{\vect{k}^2}{2m_c}-\epsilon_0, \epsilon_{M}(\vect{k})=\frac{\vect{k}^2}{2m_c}+\epsilon_0,
\end{equation}
where $m_c$ is the electron effective mass, and $\epsilon_0$ controls the indirect gap between hole and electron bands. The negative $\epsilon_0$ corresponds to the \emph{semimetal} band structure in \figref{fig: c band}, that the hole and electron bands overlap slightly in energy, which is the case for most iron-based compounds\cite{MaLu08, Singh08, Singh09} (1111-, 122-, and 111-type). The positive $\epsilon_0$ gives the \emph{semiconductor} band structure, which is also seen\cite{FeSe1, FeSe2, FeSe3} in K$_{0.4}$Fe$_{0.8}$Se. In this work, we will only focus on the negative $\epsilon_0$ semimetal case.

The chemical potential $\mu$ controls the electron doping. For this particular model, the $\Gamma$ and $M$ pockets are perfectly nested at $\mu = 0$ through the nesting vector $\vect{Q}_s$, which might not be the case in reality. The perfect nesting point here should be understood as the maximally nested case in the real material.

Admittedly, the two pocket model explicitly breaks the 4-fold rotation symmetry of Fe square lattice, and hence is not capable to describe the nematic ordering in iron-based compounds. Also the details about orbital components are neglected, so the orbital-related features are not presented in our model, such as the nodal SDW and the orbital fluctuation/ordering. Extending the two pocket model to the more realistic four pocket model and including the orbital details in the band structure will presumably enable us to incoproate the orbital physics, which has been found to be another important ingredient of iron-based compounds, into our theoretical framework. However in this work, we will first focus on the interplay between itinerant electron and local moment magnetism, and leave the orbital physics for future research. Thus as a proof-of-principle study, we will simply use the two-pocket model to describe the itinerant electron system.

The ground state of the itinerant electron system is a weakly correlated Fermi-liquid. The various Fermi-surface instabilities at low temperature are responsible for both the SDW and SC ordering in the iron-based compounds. Around $\mu = 0$ the good nesting condition gives rise to a strong SDW instability, and the itinerant electron will settle down in the SDW phase at low temperature. For larger absolute value of $\mu$ (either positive or negative), the nesting condition is destroyed, and the SDW instability fades away and gives way to the SC instability. The pairing of itinerant electron is mediated by the underlying local moment fluctuation, or the magnon, whose energy scale is much higher than the phonon glue in conventional superconductors, leading to higher SC transition temperature. Thus the high-$T_c$ SC phase appearing in the iron-based compounds upon doping is understandable.

\subsection{Spin-Liquid Description of Local Moments}
Away from the SDW phase in the iron-based superconductors, no magnetic order is observed, which suggests that the local moment can not order at low temperature by itself, unless it is assisted by coupling to a nearly nested itinerant Fermi surface with strong SDW instability. The local moment could remain disordered at low temperature, if the exchange coupling $J_{ij}$ in \eqnref{eq: H_b MM} is strongly frustrated, or if the moment amplitude fluctuates due to the relative small (orbital-selective) Mott gap\cite{Moon10,NLWang12,YYWang12} ($\sim 0.6$eV), both tending to stabilize the spin-liquid\cite{Wen91,ReadSachdev91} ground state. Many recent studies have revealed the possibility of the spin-liquid phase in the $J_1$-$J_2$ model.\cite{Richter08, Jiang09, Richter10, Reuther10, Jiang11, LWang11,TLi12,YangYao12} However given the robust collinear SDW ordering observed in many parent compounds of the iron-based materials, we suspect that even if the local moment ground state is disordered, it must be close to the collinear AFM ordering. Therefore we postulate that the local moments in the iron-based compounds may be approximately described by a spin-liquid state with short-ranged collinear AFM correlations.

The elementary excitations in the spin-liquid state include the \emph{spinons},\cite{ReadSachdev91, Sachdev94, Senthil04} which are spin-1/2 bosons as fractionalized\cite{Yao10} from the spin-1 local moment. The low-energy properties of the spin-liquid can be described by an effective Hamiltonian for spinons. On each site $i$, we introduce the Schwinger bosons $b_{i\sigma}$ ($\sigma=\uparrow,\downarrow$),\cite{Arovas88} such that the local moment $\vect{M}_i$ can be decomposed into $\vect{M}_i = \frac{1}{2} b_{i\sigma}^\dagger \vect{\sigma}_{\sigma\sigma'}b_{i\sigma'}$, with $\sum_\sigma b_{i\sigma}^\dagger b_{i\sigma}=2$. We choose the bosonic spinon (Schwinger boson) representation other than the fermionic representation, because the former can be easily connected to the magnetically ordered states (e.g. the collinear AFM in the context of iron-based materials) by condensation of the bosonic spinons. Each spinon carries spin-1/2, and on each Fe site, the local moment is made up of roughly $2$ spinons aligned in the same direction, such that the total spin is $M\simeq 1$  (corresponding to the magnetic moment $g\mu_BM$ with $g=2$). Such a setting can give quantitative account\cite{KouLiWeng09} of the ordered moment ($\sim0.8\mu_B$) observed in experiments.\cite{Goldman08, Kaneko08, Matan09}

The most general mean-field Hamiltonian includes both the spinon pairing and hopping on the Fe square lattice\cite{Si12}
\begin{equation}\label{eq: H_b bibj}
\begin{split}
H_{b}=&\sum_{i,j}\eta_{ij} \epsilon^{\sigma\sigma'} b_{i\sigma'}b_{j\sigma}+\chi_{ij} b_{i\sigma}^\dagger b_{j\sigma}+h.c.\\
&+\sum_{i}\lambda b_{i\sigma}^\dagger b_{i\sigma}.
\end{split}
\end{equation}
$\eta_{ij}$ is the spin-singlet pairing strength. $\chi_{ij}$ is the spin-independent hopping strength. $\lambda$ is the Lagrangian multiplier to control the spinon number. Here $\epsilon^{\sigma\sigma'}$ is the Levi-Civita symbol, such that $\epsilon^{\sigma\sigma'} b_{i\sigma'}b_{j\sigma}\equiv b_{i\downarrow}b_{j\uparrow}-b_{i\uparrow}b_{j\downarrow}$ denotes the singlet pairing of spinons.

The spinon representation has an U(1) gauge redundancy, as the local U(1) transformation $b_{i\sigma} \to e^{i\theta_i} b_{i\sigma}$ leaves the local moment $\vect{M}_{i}=M b_{i\sigma}^\dagger \vect{\sigma}_{\sigma\sigma'}b_{i\sigma'}$ (and hence all physical results) invariant. However the mean-field parameters in \eqnref{eq: H_b bibj} are not gauge invariant, and should transform as $\eta_{ij}\to e^{-i\theta_i-i\theta_j}\eta_{ij}$ and $\chi_{ij}\to e^{i\theta_i-i\theta_j}\chi_{ij}$. So the phase fluctuations of the mean-field parameters are subject to the U(1) gauge structure. To prevent the gauge fluctuation from confining the spinons, we require the simultaneous presentation of both the hoping and pairing mean-field, so as to break the U(1) gauge structure to $\mathbb{Z}_2$, and to gap the gauge fluctuations.\cite{Sachdev91, WangVishwanath, Wang10} In this case, the infrared behavior of mean-field theory \eqnref{eq: H_b bibj}  is well under control, and the spinon remains the only low energy excitation. We assume that the local moment is in such a $\mathbb{Z}_2$ spin-liquid state, qualitatively described by the spinon mean-field theory.

\begin{figure}[htbp]
\begin{center}
\includegraphics[width=0.24\textheight]{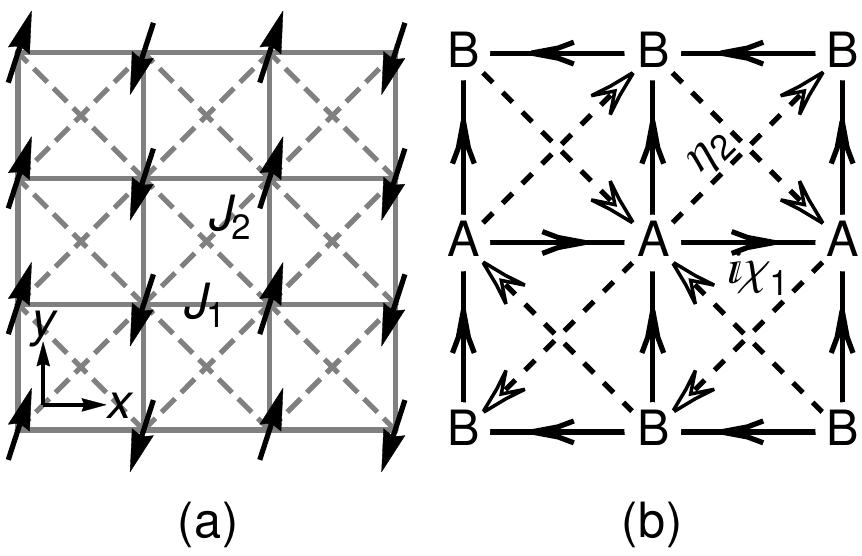}
\caption{(a) The collinear AFM configuration of local moments on the Fe lattice in a $J_1$-$J_2$-like model. (b) The arrangement of the spinon mean-field parameters. The solid arrow indicates the imaginary hopping $i\chi_1$, while the dashed arrow indicates the spin-singlet pairing $\eta_2$. Each spinon unit-cell contains two sites, labeled by $A$ and $B$ respectively.}
\label{fig: J1J2}
\end{center}
\end{figure}

Finally we expect this spin-liquid to preserve the short-range collinear AFM correlation, whose magnetic structure is shown in \figref{fig: J1J2}(a), with the ordering momentum $\vect{Q}_s=(\pi,0)$. On the mean-field level, this state can be described by the following parameters
\begin{equation}\label{eq: MF ansatz}
\begin{split}
\eta_{i,i+\hat{x}+\hat{y}} =(-)^{i_y}\eta_2,&\quad \eta_{i,i+\hat{x}-\hat{y}} =-(-)^{i_y}\eta_2,\\
\chi_{i,i+\hat{x}}=(-)^{i_y}i\chi_1,&\quad \chi_{i,i+\hat{y}}=i\chi_1.
\end{split}
\end{equation}
Here $\hat{x}$($\hat{y}$) denotes the displacement by one lattice spacing along the $x$($y$)-direction (on the Fe square lattice). $i_y$ is the vertical ($y$) component of the lattice coordinate. $\eta_2\in\mathbb{R}$ is the next-nearest-neighboring (nnn) pairing strength, and $\chi_1\in\mathbb{R}$  the nearest-neighboring (nn) hopping strength. The configuration of the mean-field ansatz is illustrated in \figref{fig: J1J2}(b). Spin-singlet pairing $\eta_2$ favors nnn AFM and imaginary hopping $i \chi_1$ favors nn FM. Both are compatible with the collinear AFM order. According to the PSG\cite{Wen02,WangVishwanath,Wang10} analysis, \eqnref{eq: MF ansatz} is the only admissible ansatz for symmetric $\mathbb{Z}_2$ spin-liquid that includes both nnn pairing and nn hoping (for details, see Appendix \ref{app: PSG}).

The parameters $\chi_1$, $\eta_2$ are usually determined by the variational approach, i.e. projecting the mean-field ground state to the physical spin Hilbert space to construct the variational wave function, on which minimizing the spin model energy to determine the optimal parameters. But we will not follow this variational approach here. We just construct a bosonic spin-liquid by symmetry arguments, and calculate its spin excitation spectrum. Then the parameters are chosen to fit the observed INS spectrums. In our approach, the spin-liquid is merely a language to describe the disordered quantum  ground state of the local moments.

\begin{figure}[htbp]
\begin{center}
\includegraphics[width=0.2\textheight]{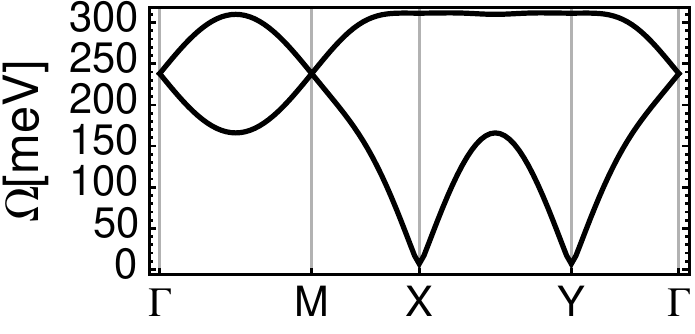}
\caption{Typical spinon band structure, with the parameters  $\lambda=238$meV, $\chi_1=36$meV, $\eta_2=34$meV. The high symmetry momentum points are defined in \figref{fig: lattice BZ}(b). }
\label{fig: b band}
\end{center}
\end{figure}

To calculate the spectrum, we switch to the momentum space, in which the mean-field Hamiltonian in \eqnref{eq: H_b bibj} takes the form
\begin{equation}
H_{b}=\frac{1}{2}\sum_{\vect{k}} \phi_{\vect{k}}^\dagger h_{b}(\vect{k}) \phi_\vect{k},
\end{equation}
with the spinon operators arranged into the Bogoliubov basis $\phi=\bigl(\begin{smallmatrix}b\\ \mathcal{T}b^\dagger \end{smallmatrix}\bigr)\otimes\bigl(\begin{smallmatrix}A\\B \end{smallmatrix}\bigr)\otimes\bigl(\begin{smallmatrix}\uparrow\\ \downarrow \end{smallmatrix}\bigr)$ following the particle-hole, sublattice and spin degrees of freedom. The time reversal operator $\mathcal{T}$ flips the spin $\bigl(\begin{smallmatrix}\uparrow\\ \downarrow \end{smallmatrix}\bigr)\to\bigl(\begin{smallmatrix}\downarrow\\ -\uparrow \end{smallmatrix}\bigr)$ and reverse the momentum $\vect{k}\to-\vect{k}$ for $b^\dagger$ operators. The matrix representation of the mean-field Hamiltonian in the Bogoliubov basis reads
\begin{equation}\label{eq: h_b 1}
\begin{split}
h_{b}(\vect{k})=&\lambda\sigma_{000}-2\chi_1(\sin k_x \sigma_{330}+\sin k_y \sigma_{310})\\
&-4\eta_2\sin k_x \sin k_y\sigma_{220},
\end{split}
\end{equation}
where $\sigma_{abc}\equiv\sigma_a\otimes\sigma_b\otimes\sigma_c$ denotes the direct product of Pauli matrices acting in the particle-hole, sublattice and spin space respectively. Its band structure is given by
\begin{equation}
\begin{split}
\Omega_{\pm} &= \sqrt{(\lambda\pm\chi_\vect{k})^2-\eta_\vect{k}^2},\\
\chi_\vect{k} &= 2\chi_1\sqrt{\sin^2k_x+\sin^2k_y},\\
\eta_\vect{k} & = 4\eta_2\sin k_x\sin k_y,
\end{split}
\end{equation} 
which includes two branches: $\Omega_+$ (upper) and $\Omega_-$ (lower) as shown in \figref{fig: b band}. The lower (upper)  branch describes the in-phase (out-of-phase) fluctuation between sublattices $A$ and $B$. 

\begin{figure}[htbp]
\begin{center}
\includegraphics[width=0.18\textheight]{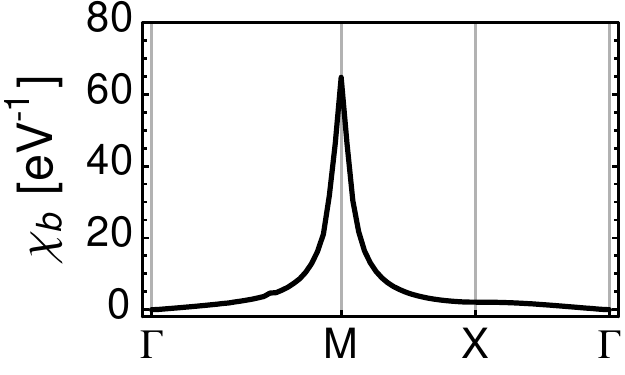}
\caption{The magnetic structural function (spin-spin correlation) of the local moment spin-liquid. The peak at $M(\pi,0)$ point indicates that the spin-liquid is close to the collinear AFM order.}
\label{fig: chi}
\end{center}
\end{figure}

We shall focus on the low-energy effect theory by projecting the Hamiltonian into the lower branch. We first perform a momentum-dependent basis rotation along the $\sigma_{020}$ direction (in the sublattice space) to transform the Hamiltonian \eqnref{eq: h_b 1} to $h_{b}(\vect{k})=\lambda\sigma_{000}+\chi_\vect{k}\sigma_{330}-\eta_\vect{k}\sigma_{220}$, which has the following block-diagonalized form
\begin{equation}
h_{b}(\vect{k})= 
\left(\begin{array}{c|cc|c}
\lambda+\chi_\vect{k} & 0 & 0 & \eta_\vect{k}\\
\hline
0 & \lambda-\chi_\vect{k} & -\eta_\vect{k} & 0\\
0 & -\eta_\vect{k} & \lambda-\chi_\vect{k} & 0\\
\hline
\eta_\vect{k} & 0 & 0 & \lambda+\chi_\vect{k}\\
\end{array}\right)\otimes\sigma_0.
\end{equation}
The inner block corresponds to the lower branch of the spinon spectrum. The Hamiltonian in this block reads $h_{b}(\vect{k})=(\lambda-\chi_\vect{k})\sigma_{00}-\eta_\vect{k}\sigma_{10}$, which acts on the truncated basis $\tilde{\phi}=\bigl(\begin{smallmatrix}b\\ \mathcal{T}b^\dagger \end{smallmatrix}\bigr)\otimes\bigl(\begin{smallmatrix}\uparrow\\ \downarrow \end{smallmatrix}\bigr)$ with the sublattice degrees of freedom projected. One may absorb $\chi_\vect{k}$ into $\lambda$ by defining $\lambda_\vect{k}=\lambda-\chi_\vect{k}$, and write
\begin{equation}
H_{b}=\sum_{\vect{k}}-\frac{1}{2}(\eta_\vect{k}\epsilon^{\sigma\sigma'} b_{-\vect{k}\sigma'}b_{\vect{k}\sigma}+h.c.)+\lambda_\vect{k}b_{\vect{k}\sigma}^\dagger b_{\vect{k}\sigma}.
\end{equation}
As shown in \figref{fig: b band}, the spinon spectrum is soften at $X,Y=(\pm\pi/2,\pi/2)$ points. Expanding the Hamiltonian around the softening points $\vect{K}=X,Y$ yields
\begin{equation}\label{eq: H_b}
\begin{split}
H_{b}=
&-\frac{1}{2}\sum_{\vect{K},\vect{k}}\eta_\vect{K}(\vect{k}) \epsilon^{\sigma\sigma'} b_{-(\vect{K}+\vect{k})\sigma'}b_{(\vect{K}+\vect{k})\sigma}+h.c.\\
&+\sum_{\vect{K},\vect{k}}\lambda_\vect{K}(\vect{k})b_{(\vect{K}+\vect{k})\sigma}^\dagger b_{(\vect{K}+\vect{k})\sigma},
\end{split}
\end{equation}
where $\vect{k}$ denotes the small momentum deviation from the softening points, and
\begin{equation}
\begin{split}
\eta_\vect{K}(\vect{k}) & \equiv\eta_{\vect{K}+\vect{k}}=4\eta_2\sin K_x \cos k_x \cos k_y,\\
\lambda_\vect{K}(\vect{k}) & \equiv\lambda_{\vect{K}+\vect{k}}= \lambda - 2\chi_1\sqrt{\cos^2k_x+\cos^2k_y}.
\end{split}
\end{equation}
Thus we have reduced the spinon lattice model \eqnref{eq: H_b bibj} to its low-energy effective model \eqnref{eq: H_b} around the band softening points. Its band structure contains two valleys, each located at a band softening point, of the dispersion
\begin{equation}
\Omega_\vect{K}(\vect{k}) = \sqrt{c_b^2\vect{k}^2+m_b^2},
\end{equation}
with the spinon velocity $c_b= (4\eta _2 (4 \eta _2+\sqrt{2} \chi _1))^{1/2}$ and the spinon gap $m_b=((\lambda-2\sqrt{2}\chi_1)^2-16\eta_2^2)^{1/2}$. The valleys are connected by the momentum $\vect{Q}_s=(\pi,0)$. The inter-valley scattering of the spinons leads to strong magnetic fluctuations at $\vect{Q}_s$ for the local moment, as shown in \figref{fig: chi}.

Therefore the spinon mean-field Hamiltonian \eqnref{eq: H_b} does describe a fully gapped $\mathbb{Z}_2$ spin-liquid close to the collinear AFM order. Such a spin-liquid description can simultaneously capture the long-range magnetic disorder and the short-range collinear AFM correlation of the local moments. It could be more appropriate if the local moment is not well-quantized, such that the confinement force between the spinons is weaker. This is indeed the case in the iron-based compounds, where the orbital-selective Mott gap is not very large ($\sim0.6$eV),\cite{YYWang12} which may be sufficient to suppress the charge fluctuation, but not large enough to precisely quantize the local moment. This leads to the ring-exchange interaction among the local moments, which tends to stabilize\cite{Yao10} the deconfined spin-liquid state in the spin-1 systems. So it could be easier for the local moment to fractionalize into deconfined spinons in the intermediate correlated systems such as the iron-based compounds, which supports the spin-liquid description.

\subsection{Hund's Rule Coupling}
Using the spinon representation, the Hund's rule interaction $H_\text{int}$ in \eqnref{eq: H_int SM} can be express as follows
\begin{equation}
H_{cb} = -J_0\sum_{i} c_{i\sigma_c}^\dagger \vect{\sigma}_{\sigma_c\sigma'_c}c_{i\sigma'_c} \cdot b_{i\sigma_b}^\dagger \vect{\sigma}_{\sigma_b\sigma'_b}b_{i\sigma'_b},
\end{equation}
where $J_0\equiv J_{H}M/2$ is the effective coupling strength between itinerant electrons and spinons. $\sigma_c,\sigma'_c$ and $\sigma_b,\sigma'_b$ label their spins respectively, which are summed over implicitly.

In the momentum space, the Hamiltonian reads
\begin{equation}\label{eq: Hcb}
\begin{split}
H_{cb} = -J_0 & \sum_{\vect{K}_b,\vect{k}_b}\sum_{\vect{K}_c,\vect{k}_c}\sum_{\vect{Q},\vect{q}} \\
&c_{(\vect{K}_c-\vect{Q}+\vect{k}_c-\vect{q})\sigma_c}^\dagger \vect{\sigma}_{\sigma_c\sigma'_c}c_{(\vect{K}_c+\vect{k}_c)\sigma'_c}\\
&\cdot b_{(\vect{K}_b+\vect{Q}+\vect{k}_b+\vect{q})\sigma_b}^\dagger \vect{\sigma}_{\sigma_b\sigma'_b}b_{(\vect{K}_b+\vect{k}_b)\sigma'_b}.
\end{split}
\end{equation}
$\vect{K}_c= \Gamma, M$ labels the two pockets of itinerant electrons. $\vect{K}_b =X,Y$ labels the two band softening points of spinons. The momentum transfer $\vect{Q} =(0,0)$ or $(\pi,0)$ corresponds to either of the two scattering channels: the intro-pocket scattering $\vect{Q}=(0,0)$ in which the itinerant electrons are scattering within the same pocket and the spinons are scattered around the same softening point; and the inter-pocket scattering $\vect{Q}=(\pi,0)$ in which the itinerant electrons are transferred form one pocket to another and meanwhile the spinons are transferred between two softening points to conserve the momentum (see \figref{fig: lattice BZ}). Both the SDW and SC instabilities develop mainly through the inter-pocket scattering channel. While both scattering channels are responsible for the hump-dip feature in the normal phase.

The Hund's rule interaction between the electrons and the spinons can be treated perturbatively. We will show that this interaction gives rise to two kinds of low-energy collective modes: the magon mode and the composite fermion mode. In the following, we will conclude the two-fluid model in the field theory language, and use random phase approximation (RPA) approach to analyze the collective mode beyond the mean-field theory.

\section{Field Theory Discription}
\subsection{Model Action}

It is more convenient to work in the field theory (path integral) formalism than in the above Hamiltonian formalism. So we introduce the electron and spinon field in the imaginary frequency ($i\omega$) domain
\begin{equation}
c_k = \int\mathrm{d}\tau e^{i\omega\tau} c_{\vect{k}},b_k = \int\mathrm{d}\tau e^{i\omega\tau} b_{\vect{k}}.
\end{equation}
$k=(i\omega,\vect{k})$ denotes the energy-momentum vector. Following the order of particle-hole, pocket (valley), and spin degrees of freedom, we arrange the electron and the spinon fields as
\begin{equation}
\begin{split}
\psi_{c}(k) &=\left(
\begin{smallmatrix}
 c_k \\
 \mathcal{T} c_{k}^{\dagger }
\end{smallmatrix}
\right)\otimes \left(
\begin{smallmatrix}
 \Gamma  \\
 M
\end{smallmatrix}
\right)\otimes \left(
\begin{smallmatrix}
 \uparrow  \\
 \downarrow 
\end{smallmatrix}
\right),\\
\psi_{b}(k) &=\left(
\begin{smallmatrix}
 b_k \\
 \mathcal{T} b_{k}^{\dagger }
\end{smallmatrix}
\right)\otimes \left(
\begin{smallmatrix}
 X\\
 Y
\end{smallmatrix}
\right)\otimes \left(
\begin{smallmatrix}
 \uparrow  \\
 \downarrow 
\end{smallmatrix}
\right).
\end{split}
\end{equation}
$\mathcal{T}$ stands for the time reversal operator, which flips the momentum $\mathcal{T} k\to -k$, and acts on the spin degrees of freedom as $\mathcal{T}(\uparrow,\downarrow)\to(\downarrow,-\uparrow)$. $\Gamma, M$ are the central momenta of the two electron pockets. $X, Y$ are the softening momenta of the spinons.

The partition function of the two-fluid model reads $Z=\int\mathcal{D}[\psi_c,\psi_b]e^{-S}$, with the action $S=S_c+S_b+S_{cb}$,
\begin{equation}
\begin{split}
S_c &= -\sum_k \psi_c^\dagger(k)G_{c(0)}(k)^{-1}\psi_c(k),\\
S_b &= -\sum_k \psi_b^\dagger(k)G_{b(0)}(k)^{-1}\psi_b(k),\\
S_{cb} &= -\sum_{k,k',q} \psi_c^\dagger(k+q)\psi_b^\dagger(k'-q)X \psi_b(k')\psi_c(k),\\
\end{split}
\end{equation}
where the matrix $X=(J_0/4)\sum_{a=0,1}\sum_{i=1,2,3}\sigma_{0ai}\otimes\sigma_{3ai}$ in the $\psi_b\otimes\psi_c$ basis. According to \eqnref{eq: H_cc} and \eqnref{eq: H_b}, the bare propagators are
\begin{equation}\label{eq: G0}
\begin{split}
G_{c(0)}(k) &=2(i\omega\sigma_{000}+\mu\sigma_{300}+\epsilon(\vect{k})\sigma_{330})^{-1},\\
G_{b(0)}(k) &=2(i\omega\sigma_{300}+\lambda(\vect{k})\sigma_{000}+\eta(\vect{k})\sigma_{130})^{-1}.
\end{split}
\end{equation}
The short-handed notion $\sigma_{abc\cdots}=\sigma_a\otimes\sigma_b\otimes\sigma_c\otimes\cdots$ ($a,b,c=0,1,2,3$) denotes the direct product of Pauli matrices. $\mu$ is the itinerant electron chemical potential, and the two-pocket band structure is described by
\begin{equation}
\epsilon(\vect{k})=\frac{\vect{k}^2}{2m_c}+\epsilon_0,
\end{equation}
parameterized by $m_c$ and $\epsilon_0<0$. The spinon band structure around the band softening point is modeled by
\begin{equation}
\begin{split}
\eta(\vect{k}) &= 4\eta_2 \cos k_x \cos k_y,\\
\lambda(\vect{k}) &= \lambda - 2\chi_1\sqrt{\cos^2 k_x+\cos^2 k_y},
\end{split}
\end{equation}
which is controlled by the parameters $\lambda$, $\chi_1$ and $\eta_2$.

The parameters can be estimated from the experimental input. By comparing itinerant electron band structure with the LDA calculation or the ARPES observation, we may take $m_c=0.83$eV$^{-1}$ and $\epsilon_0= -30$meV, which produce the reasonable pocket size and depth. For spinons, we choose $\chi_1=36$meV $\eta_2 = 34$meV, $\lambda = 238$meV, so that the gap $m_b=7$meV and velocity $c_b=159$meV can lead to a magnon spectrum comparable with the INS observation. The residual Hund's rule coupling $J_0$ is the only free parameter in our model. It may vary in different phases depending on the RG flow from the high-energy regime.

\subsection{Decomposition of the Coupling}

\begin{figure}[htbp]
\begin{center}
\includegraphics[width=0.3\textheight]{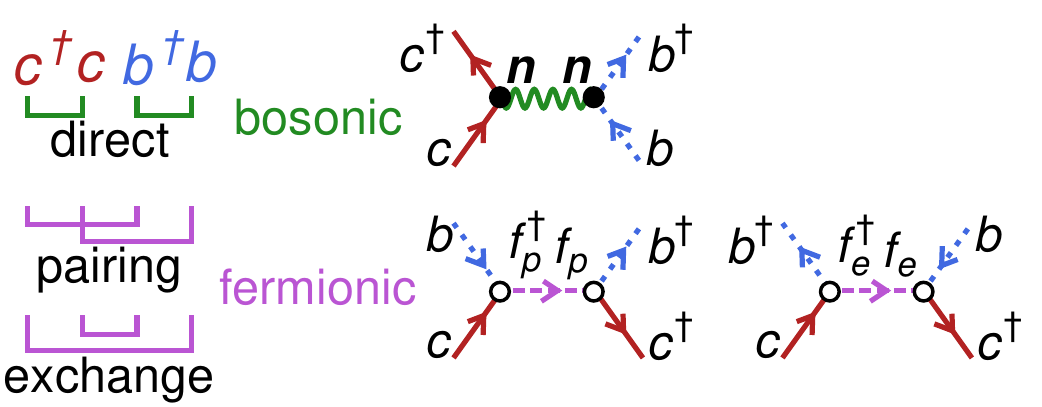}
\caption{Decomposition of the electron-spinon coupling in the direct, pairing and exchange channels. The direct channel leads to bosonic magnon $\vect{n}$. The pairing and exchange channels lead to the fermionic composite modes $f_p$ and $f_e$ respectively.}
\label{fig: HS}
\end{center}
\end{figure}

To handle the electron-spinon coupling term $S_{cb}$ (or $H_{cb}$) in the two-fluid model,
\begin{equation}
H_{cb}\simeq-J_0 c^\dagger\vect{\sigma}c\cdot b^\dagger\vect{\sigma}b,
\end{equation}
we resort to the Hubbard-Stratonovich decomposition. There are three possible decomposition channels: direct, pairing and exchange. We use the solid line $\diag{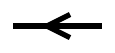}{8pt}=\langle \psi_c\psi_c^\dagger\rangle$ to denote the electron propagator, and the dotted line $\diag{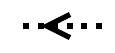}{8pt}=\langle\psi_b\psi_b^\dagger\rangle$ to denote the spinon propagator. The decomposition schemes of electron-spinon coupling are illustrated in \figref{fig: HS}.

Here we briefly sketch out the decomposition schemes, and the details will be provided in the next two sections. (i) The direct channel is a bosonic channel, $H_{cb} \to  H_{ncb}+H_{n}$,
\begin{equation}
\begin{split}
H_{ncb} &\simeq-J_0(\vect{n}_b\cdot c^\dagger\vect{\sigma}c+\vect{n}_c\cdot b^\dagger\vect{\sigma}b), \\
H_{n} &\simeq J_0 \vect{n}_b\cdot\vect{n}_c,
\end{split}
\end{equation}
which describes the electron and the spinon interact with each other by exchanging the magnon $\vect{n}_{c,b}$ (bosonic spin-wave excitations). In general, we use the wavy line $\diag{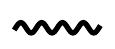}{8pt}=\langle \vect{n}\vect{n}\rangle$ to denote the magnon propagator.  The bosonic channel is responsible for the low-energy spin dynamics. (ii) The pairing and exchange channels are fermionic channels, $H_{cb} \to H_{fcb}+H_{f}$,
\begin{equation}
\begin{split}
H_{fcb} &\simeq \frac{J_0}{2}(f_p^\dagger c b+f_e^\dagger c b^\dagger+h.c.),\\
H_{f} &\simeq \frac{J_0}{2}(f_p^\dagger f_p+f_e^\dagger f_e),
\end{split}
\end{equation}
in which the electron and the spinon combine into the composite fermion $f_p$ (for the pairing channel) or $f_e$ (for the exchange channel). We may arrange the composite fermions in the field $\psi_f$ (to be defined in \eqnref{eq: psi_f} later), and denote its propagator by the dashed line $\diag{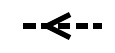}{8pt}=\langle \psi_f\psi_f^\dagger\rangle$. The fermionic channels is responsible for the low-energy charge dynamics. We will discuss the two channels (i) and (ii) respectively in the following two parts.

\section{Spin Dynamics}
\subsection{Bosonic Decomposition}
To rudy the spin dynamics of the two-fluid model, we start from the decomposition of the electron-spinon interation $H_{cb}$ in the bosonic channel (magnon channel). Let $\vect{s}_{c(b)}$ be the electron (spinon) spin operator, \eqnref{eq: Hcb} can be written as
\begin{equation}
\begin{split}
H_{cb} &= -J_0\sum_{\vect{Q},\vect{q}} \vect{s}_{c(-\vect{Q}-\vect{q})}\cdot\vect{s}_{b(\vect{Q}+\vect{q})},\\
\vect{s}_{c(\vect{Q}+\vect{q})}& \equiv \sum_{\vect{K}_c,\vect{k}_c}c_{(\vect{K}_c+\vect{Q}+\vect{k}_c+\vect{q})\sigma_c}^\dagger \vect{\sigma}_{\sigma_c\sigma'_c}c_{(\vect{K}_c+\vect{k}_c)\sigma'_c},\\
\vect{s}_{b(\vect{Q}+\vect{q})}& \equiv \sum_{\vect{K}_b,\vect{k}_b} b_{(\vect{K}_b+\vect{Q}+\vect{k}_b+\vect{q})\sigma_b}^\dagger \vect{\sigma}_{\sigma_b\sigma'_b}b_{(\vect{K}_b+\vect{k}_b)\sigma'_b},
\end{split}
\end{equation}
It is natural to introduce the field $\vect{n}_{c,b}$ that couples to $\vect{s}_{b,c}$ respectively. Then the electron-spinon coupling can be decomposed in the direct channel as $H_{cb} \to H_{ncb}+H_{n}$
\begin{equation}
\begin{split}
H_{ncb} = -J_0\sum_{\vect{Q},\vect{q}} &\vect{n}_{b(-\vect{Q}-\vect{q})}\cdot \vect{s}_{c(\vect{Q}+\vect{q})}\\
&+\vect{n}_{c(-\vect{Q}-\vect{q})}\cdot \vect{s}_{b(\vect{Q}+\vect{q})},
\end{split}
\end{equation}
\begin{equation}
H_{n} = J_0 \sum_{\vect{Q},\vect{q}} \vect{n}_{b(-\vect{Q}-\vect{q})}\cdot \vect{n}_{c(\vect{Q}+\vect{q})}.
\end{equation}
Integrate out $\vect{n}_{c,b}$ field will restore the Hamiltonian $H_{cb}$. $\vect{n}_c$ ($\vect{n}_b$) represents the collective spin fluctuation of itinerant electron (local moment). 

To simplify the notation, we arrange the different $\vect{Q}$ components of the magnon field $\vect{n}$ into a combined field $n(q)$. Take $n_b$ field for example ($n_c$ is defined similarly by changing the label $b\to c$)
\begin{equation}
n_b(q)=\int\mathrm{d}\tau e^{i\nu\tau}\left(\begin{matrix}\vect{n}_{b\vect{q}}\\ \vect{n}_{b(\vect{Q}_s+\vect{q})}\end{matrix}\right),
\end{equation}
where $q=(i\nu,\vect{q})$ denotes the energy-momentum vector, and $\vect{Q}_s\equiv(\pi,0)$ is the intra-valley momentum transfer. Then $e^{-S_{cb}} = \int\mathcal{D}[n_c,n_b]e^{-S_n-S_{ncb}}$,
\begin{equation}\label{eq: bosonic model}
\begin{split}
S_n &= J_0\sum_q n_b(-q)n_c(q),\\
S_{ncb} &= -J_0\sum_{q,k} (n_b(-q)\psi_c^\dagger(k+q) s_c \psi_c(k)\\
&\phantom{=-J_0\sum_{q,k}}+n_c(-q)\psi_b^\dagger(k+q) s_b \psi_b(k)),
\end{split}
\end{equation}
where the matrix representation for the electron spin operators are $s_c=(\sigma_{00i},\sigma_{01i})/2$, and that for the spinon spin operators are $s_b=(\sigma_{30i},\sigma_{31i})/2$, with $i=1,2,3$.

\subsection{Magnon}

The magnon (spin-wave) excitation can emerge from the collective dynamics of the spinon. This magnon mode, in terms of the fluctuation of $n$ field, can then couple with the itinerant electron via $S_{ncb}$ term, and drives the electron into the SDW or SC phase depending on the electron doping.

\begin{figure}[htbp]
\begin{center}
\includegraphics[width=0.36\textheight]{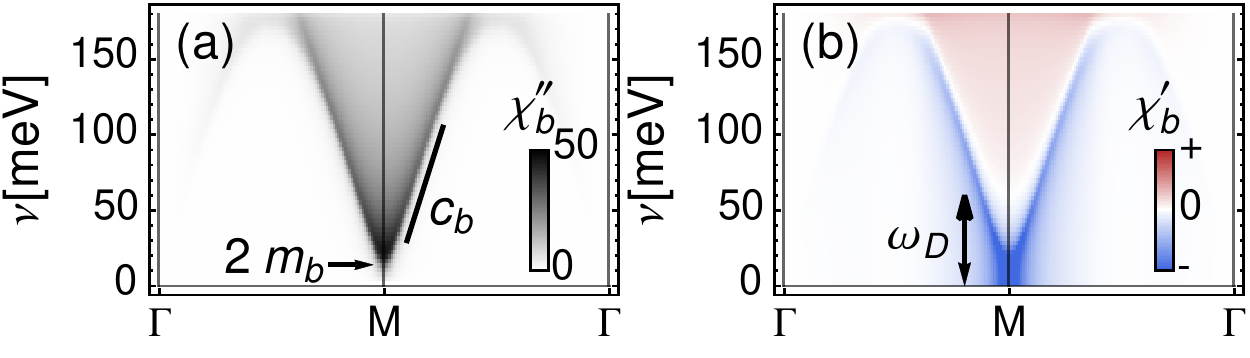}
\caption{The local moment spin susceptibility $\chi_b$. (a) $\chi_b''$ the spectral function (the imaginary part) (b) $\chi_b'$ the real part. In (a), the quasi-sharp spectral peak (dark shaded edges) emerges at the bottom of the spinon continuum (the shaded region). In (b), $\chi_b'$ is negatively large below the magnon dispersion, which is of a typical energy $\omega_D$ (see \eqnref{eq: Vq} in text).}
\label{fig: magnon}
\end{center}
\end{figure}

To demonstrate the magnon mode, we first integrate out the spinon field $\psi_b$ in \eqnref{eq: bosonic model}, and obtain the effective action for the magnon field $n_c$
\begin{equation}\label{eq: Sn with bc}
S_n=\sum_q \Big(J_0 n_b(-q)n_c(q)+\frac{1}{2} n_c(-q)\Sigma_{n_c}(q)n_c(q)\Big).
\end{equation}
On the one loop level, the self-energy $\Sigma_{n_c}(q)=J_0^2\chi_b(q)$ is proportional to the spin susceptibility $\chi_b$ of spinons, defined by the following bubble diagram
\begin{equation}\label{eq: chi b}
\chi_{b}(q)=-\diag{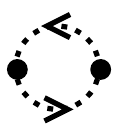}{20pt}=-\sum_k \Tr G_{b(0)}(k+q)s_b^\dagger G_{b(0)}(k)s_b,
\end{equation}
Here $s_b=\sigma_{30i}/2$ for the $\vect{Q}=(0,0)$ component, and $s_b=\sigma_{31i}/2$ for the $\vect{Q}=(\pi,0)$ component. Further integrating out the $n_c$ field in \eqnref{eq: Sn with bc}, we obtain the effective theory for the $n_b$ field,
\begin{equation}
S_n=-\frac{1}{2}\sum_q n_b(-q)\chi_b(q)^{-1}n_b(q),
\end{equation}
$\chi_b$ becomes the propagator of the local moment magnon field $n_b$. Let $\chi_{b}''(\nu,\vect{q})\equiv-2\Im \chi_{b}(q)|_{i\nu\to\nu+i0_+}$ be the spectral function of $\chi_b$, then according to \eqnref{eq: G0} and \eqnref{eq: chi b}
\begin{equation}
\begin{split}
\chi_{b}''(\nu,\vect{q})=\sum_{\vect{k}}&\left(\frac{\lambda(\vect{k})\lambda(\vect{k}+\vect{q})}{\Omega(\vect{k})\Omega(\vect{k}+\vect{q})}-\frac{\eta(\vect{k})\eta(\vect{k}+\vect{q})}{\Omega(\vect{k})\Omega(\vect{k}+\vect{q})}-1\right)\\
&\;\delta[\nu-\Omega(\vect{k})-\Omega(\vect{k}+\vect{q})]
\end{split}
\end{equation}
where $\Omega(\vect{k})=\sqrt{\lambda(\vect{k})^2-\eta(\vect{k})^2}$ is the spinon dispersion. The result is shown in \figref{fig: magnon}.

The local moment magnon spectrum is a convolution of the spinon spectrum, and so one may expect a pair-spinon continuum. However the continuum is not featureless. Quasi-sharp peak appears at the bottom of the continuum (and decays as $\chi_{b}''\sim\nu^{-1}$), which is identified as the (quasi-)magnon mode. Its quasi-sharp spectral weight signifies the close-to-condensation spinon dispersion, where the spinon spectral weight is greatly enhanced by the spinon pairing at the bottom of the spinon dispersion. This leads to the enhancement along the bottom boundary of the pair-spinon continuum, which results in the quasi-peak feature that can be identified as the magnon mode.

The magnon mode is the strongest around the ordering momentum $\vect{Q}_s=(\pi,0)$, near which the magnon follows the relativistic dispersion $\nu\simeq\sqrt{c_n^2 \vect{q}^2+m_n^2}$, where $\vect{q}$ denotes the momentum deviation from $\vect{Q}_s$. The magnon velocity $c_n$ is roughly the same as the spinon velocity $c_b$ at low energy. The magnon gap $m_n$ is twice of the spinon gap $m_b$, i.e. $m_n=2m_b\simeq 14$meV, because it requires at least $2m_b$ energy to excite two spinons, each at $X, Y$ respectively, to generate such a magnon. A relatively small (but finite) magnon mass corresponds to a disordered local moment background closed to the collinear AFM ordering. The correlation length $\xi$ can be estimated from $\xi\sim c_n/m_n$.

So now the model action is reduced to $S=S_c+S_n+S_{nc}$,
\begin{equation}\label{eq: magnon model}
\begin{split}
S_c &= -\sum_k \psi_c^\dagger(k)G_{c(0)}(k)^{-1}\psi_c(k),\\
S_n &= -\frac{1}{2}\sum_q n_b(-q)\chi_{b}(q)^{-1}n_b(q),\\
S_{nc} &=-J_0\sum_{q,k}n_b(-q)\psi_c^\dagger(k+q)s_c\psi_c(k),
\end{split}
\end{equation}
which describes the itinerant electron $\psi_c$ and the local moment fluctuation $n_b$ coupling together via effective Hund's rule interaction $S_{nc}$. This simple model has been studied in Ref.\,\onlinecite{YouPRB11}, and was found to give rise to both the SDW and the SC phase in the iron-based material under the same frame-work. In the following, we will briefly review its main results.

\subsection{SDW Phase}

Because the magnon mode is the strongest around the collinear ordering momentum $\vect{Q}_s=(\pi,0)$, which is also the nesting momentum between the hole and electron pockets, so the itinerant electron is strongly scattered between the pockets by the magnon, which leads to SDW instability under good nesting condition. The SDW ordering comes from a joint effort of both the electron and the spinon. The itinerant electron has the tendency to develop the collinear SDW order and open the SDW gap to lower its Fermi energy. While the spinon tends to close its mass gap and to condense into the collinear AFM state, so as to reduce its boson energy. The resonant scattering between electron and spinon mutually enhance the ordering tendency for both, and leads to the joint magnetic ordering.

\begin{figure}[htbp]
\begin{center}
\includegraphics[width=0.32\textheight]{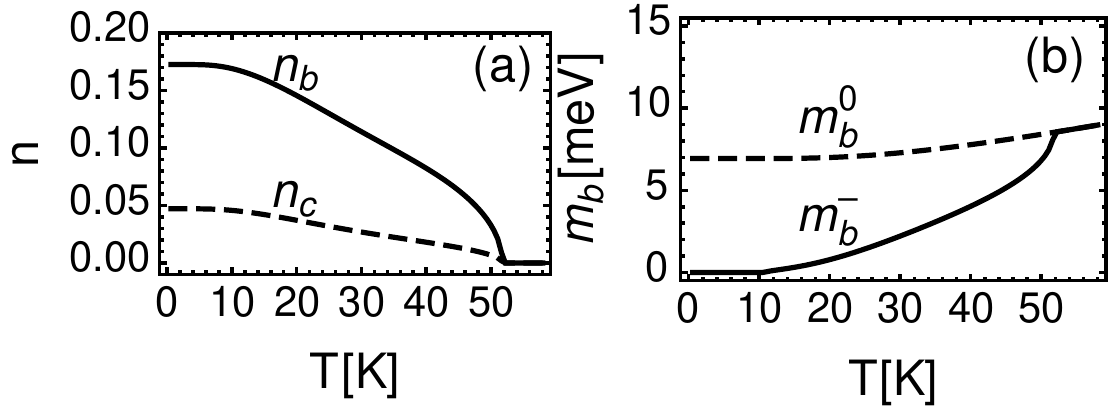}
\caption{(a) The SDW order $n_{c,b}$ v.s. temperature $T$, calculated for $\mu=0$ with $J_0=16$meV. The electron and the spinon orders simultaneously below the SDW transition temperature. (b) The spinon gap $m_b^-$ ($m_b^0$) with (without) the electron-spinon coupling $J_0=16$meV ($J_0=0$). By coupling to the electron, the spinon energy spectrum is softened (but remaining gapped) in the SDW phase, and only becomes gapless at low enough temperature.}
\label{fig: SDW}
\end{center}
\end{figure}

The joint ordering can be formulated by the following Dynson equations
\begin{equation}
\diag{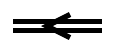}{8pt}=\diag{fig_Gc0}{8pt}+\diag{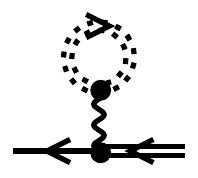}{28pt},
\diag{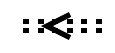}{8pt}=\diag{fig_Gb0}{8pt}+\diag{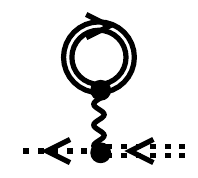}{28pt},
\end{equation}
or
\begin{equation}\label{eq: SDW Dynson}
\begin{split}
G_c(k)^{-1} &= G_{c(0)}(k)^{-1} - J_0 s_c\sum_{k'}\Tr G_b(k') s_b,\\
G_b(k)^{-1} &= G_{b(0)}(k)^{-1} + J_0 s_b\sum_{k'}\Tr G_c(k') s_c.
\end{split}
\end{equation}
which describe the mutual self-energy corrections in the direct channel of the magnetic interaction between the electron and the spinon. Given $G_{b(0)}$ and $G_{c(0)}$ in \eqnref{eq: G0}, the dressed propagator $G_b$ and $G_c$ can be solved based on \eqnref{eq: SDW Dynson} self-consistently. The collinear SDW order develops in the $\vect{Q}=\vect{Q}_s$ channel. We focus on this channel, taking $s_b = \sigma_{31i}/2, s_c =\sigma_{01i}/2$, and introduce the SDW order parameters $n_b=-\sum_k\Tr G_b(k)s_b$ (spinon magnetic order), $n_c=\sum_k\Tr G_c(k)s_c$ (electron magnetic order). The mean-field equations of the order parameters follow from the Dynson equations \eqnref{eq: SDW Dynson},
\begin{equation}
\begin{split}
n_b &= -\frac{1}{2}\sum_{s=\pm}\sum_{\vect{k}}\frac{J_0n_c+s\lambda(\vect{k})}{2\Omega_s(\vect{k})}\coth\frac{\beta\Omega_s(\vect{k})}{2},\\
n_c &= \frac{1}{2}\sum_{s=\pm}\sum_{\vect{k}}\frac{J_0 n_b}{2s E(\vect{k})}\tanh\frac{\beta(\mu+sE(\vect{k}))}{2},
\end{split}
\end{equation}
where $\Omega_s(\vect{k})=\sqrt{(\lambda(\vect{k})+sJ_0 n_c)^2-\eta(\vect{k})^2}$, and $E(\vect{k})=\sqrt{\epsilon(\vect{k})^2+(J_0n_b)^2}$. When solving the mean-field equations, we must also adjust the spinon chemical potential $\lambda$, such that the average spinon density $\rho_b=-\sum_k \Tr G_b(k)$ is kept at a fixed level $\rho_{b0}$:
\begin{equation}
\rho_{b0}=\frac{1}{2}\sum_{s=\pm}\sum_{\vect{k}}\frac{\lambda(\vect{k})+sJ_0n_c}{2\Omega_s(\vect{k})}\coth\frac{\beta\Omega_s(\vect{k})}{2}.
\end{equation}
From the solution, we can obtain the electron (spinon) SDW order $n_c$ ($n_b$) and the spinon mass gap $m_b^-=\Omega_{-}(\vect{k}=0)$. Their temperature dependences are shown in \figref{fig: SDW}. 

As the temperature drops below the SDW transition temperature, the SDW gap opens up in the electron spectrum. The electron SDW ordering $n_c=\langle\psi_c^\dagger\sigma_{01i}\psi_c\rangle$ can be sensed by the spinon via the effective Hund's rule coupling. This causes the Zeeman splitting $\pm J_0 n_c$ in the spinon spectrum, leading to the spinon SDW ordering $n_b=\langle\psi_b^\dagger\sigma_{31i}\psi_b\rangle$, which in turn supports the electron SDW gap $J_0 n_b$. This positive feedback between the electron and the spinon stabilizes the SDW ordering against thermal fluctuations, such that even a weak SDW order of the electrons can be sustained up to a rather high temperature (see \figref{fig: SDW}(a)). 

\subsection{SC Phase}
Analogous to the phonon mediated BCS pairing in the conventional superconductor, we proposed\cite{YouPRB11} that the iron-based superconductor originated from the magnon mediated pairing. The itinerant electron provides the charge carrier, and the local moment fluctuations serves as the pairing glue. At low temperature, the magnon mediated effective interaction could cause the Cooper instability of the Fermi surface of the itinerant electron, leading to the SC ordering.

We start from the effective interaction mediated by the magnon. Integrating out $n_b$ in \eqnref{eq: magnon model}, one finds $e^{-S_\text{int}}=\int\mathcal{D}[n_b]e^{-S_n-S_{nc}}$ with
\begin{equation}
S_\text{int} =\frac{1}{2}\sum_{k,k',p}\psi_c^\dagger(k+p)\psi_c^\dagger(-k)\Gamma(k-k')\psi_c(-k')\psi_c(k'+p),
\end{equation}
where the vertex function is given by
\begin{equation}
\begin{split}
\Gamma(q) &= J_0^2 \chi_{b}(q) s_c\otimes s_c\\
&= J_0^2 (\chi_{b0} \sigma _{00i}\otimes\sigma_{00i}+\chi_{b1} \sigma _{01i}\otimes\sigma _{01i}),
\end{split}
\end{equation}
in which $i$ is implicitly sumed over $i=1,2,3$. $\chi_{b0}$ is the spinon susceptibility around $\vect{Q}=(0,0)$, while $\chi_{b1}$ is that around $\vect{Q}=(\pi,0)$. As for the spin-liquid state closed to collinear AFM ordering, $\chi_b(q)$ is the most prominent around the ordering momentum $\vect{Q}=(\pi,0)$, i.e. $|\chi_{b1}|\gg |\chi_{b0}|$, so we may focus on the $\chi_{b1}\simeq-m_n^{-1}$ term, and take
\begin{equation}
\Gamma(q)\simeq -\frac{J_0^2}{m_n}\sigma_{01i}\otimes\sigma_{01i}.
\end{equation} 

Among the 64 eigenvalues of the matrix $\Gamma(q)$, the lowest (most negative) ones are 8-fold degenerate, corresponding to the 8 strongest ordering tendencies driven by the effective interaction, in which 4 of them are enumerated in \tabref{tab: instability}, and the other 4 are the charge-conjugate order of them. These orders are consistent with the more detailed analysis in Ref.\,\onlinecite{Zhai09}. Each order has an order parameter in the form of $\psi_c^\dagger u \psi_c$, with the matrix representation $u$ listed in \tabref{tab: instability}. 

\begin{table}[htdp]
\begin{center}
\begin{tabular}{c|c|c}
Order & Order Parameter & Matrix $u$\\
\hline
SC (intra) &
$c_{\Gamma\uparrow}c_{\Gamma\downarrow}-c_{\Gamma\downarrow}c_{\Gamma\uparrow}-c_{M\uparrow}c_{M\downarrow}+c_{M\downarrow}c_{M\uparrow}$ &
$\sigma_{130},\sigma_{230}$\\
SC (inter) &
$c_{\Gamma\uparrow}c_{M\downarrow}-c_{\Gamma\downarrow}c_{M\uparrow}-c_{M\uparrow}c_{\Gamma\downarrow}+c_{M\downarrow}c_{\Gamma\uparrow}$ &
$\sigma_{120},\sigma_{220}$\\
PI &
$c_{\Gamma\uparrow}^\dagger c_{\Gamma\uparrow}+c_{\Gamma\downarrow}^\dagger c_{\Gamma\downarrow} -c_{M\uparrow}^\dagger c_{M\uparrow}-c_{M\downarrow}^\dagger c_{M\downarrow}$ &
$\sigma_{030},\sigma_{330}$\\
CDW &
$c_{\Gamma\uparrow}^\dagger c_{M\uparrow}+c_{\Gamma\downarrow}^\dagger c_{M\downarrow} -c_{M\uparrow}^\dagger c_{\Gamma\uparrow}-c_{M\downarrow}^\dagger c_{\Gamma\downarrow}$ &
$\sigma_{020},\sigma_{320}$
\end{tabular}
\end{center}
\caption{Four strongest ordering tendencies driven by the magnon mediated effective interaction. The orders are: SC - superconductivity (the label indicating the intra/inter-pocket channel), PI - Pomeranchuk instability, and CDW - charge density wave. The matrix form of the order parameter is given in the last column. For example, the intra-pocket SC order can be written as a linear combination of $\psi_c^\dagger \sigma_{130}\psi_c$ and $\psi_c^\dagger \sigma_{230}\psi_c$.}
\label{tab: instability}
\end{table}

To examine which ordering tendency has the strongest Fermi surface instability, we consider the itinerant electron propagator with the order parameter (extended from the bare propagator $G_{c(0)}$ in \eqnref{eq: G0})
\begin{equation}
G_{c}=2(i\omega\sigma_{000}+\mu\sigma_{300}+\epsilon\sigma_{330}-\Delta u)^{-1},
\end{equation}
where $\Delta$ denotes the self-energy of the order under consideration. To fully gap up the Fermi surface, the order parameter matrix $u$ must anti-commute with both $\sigma_{300}$ and $\sigma_{330}$. Only the intra-pocket SC ordering ($u=\sigma_{130}, \sigma_{230}$) satisfies this condition, and enjoys the perfect nesting instability. Other orders like the inter-pocket SC and the PI only have the Stoner instability, and the CDW suffers from the bad nesting (for $\mu\neq0$). In conclusion, the magnon mediated interaction naturally selects out the spin-singlet intra-pocket pairing with $s_\pm$-wave symmetry (as $\cos k_x\cos k_y$), that the pairing order parameter changes sign between the hole and electron pockets while remains nodeless around each pocket. This is consistent with the mainstream understanding of the pairing symmetry of iron-based superconductors as reviewed in Ref.\,\onlinecite{Mazin09}.

So we focus on the spin-singlet $s_\pm$-wave SC described by the order parameter $\psi_c^\dagger \sigma_{130}\psi_c$ (assuming real pairing). If the $\Gamma$ and $M$ bands are not symmetric to each other with respect to the Fermi energy (when $\mu\neq0$), the two pockets will have different pairing strength, then the ordinary $s^{++}$-wave pairing component $\psi_c^\dagger\sigma_{100}\psi_c$ will be induced from the $s_\pm$-wave pairing. So it is more convenient to consider the pairing on both pockets respectively, and define their pairing operators as $\eta_{\Gamma(M)}(k)=\frac{1}{2}\psi_c^\dagger(k)(\sigma_{100}\pm\sigma_{130})\psi_c(k)$, which may be arranged into a vector $\eta=(\eta_\Gamma,\eta_M)^\intercal$. The effective pairing interaction scatters the Cooper pair between pockets,
\begin{equation}
S_\text{pair}=\sum_{k,k'}V(k-k')\eta(k)\sigma_1\eta(k'),
\end{equation} 
where $V(q)= - 3 J_0^2 \chi_{b1}(q)$ is the pairing potential. $V(q)$ is large below the magnon dispersion (in the gap below the spinon continuum), as shown in \figref{fig: magnon}(b). Thus we introduce the \emph{magnon Debye energy} $\omega_D$ as the typical magnon energy, and approximate $V(q)$ as
\begin{equation}\label{eq: Vq}
V(q)=\left\{\begin{array}{ll}V(\vect{q})&\nu<\omega_D\\0 & \nu>\omega_D\end{array}\right.,
\end{equation}
where $V(\vect{q})\simeq 3 J_0^2(c_n^2\vect{q}^2+m_n^2)^{-1/2}$ is taken from the $\nu=0$ component of $V(q)$, which is a positive definite (repulsive) potential. This allows us to establish the Hamiltonian of pairing interaction,
\begin{equation}\label{eq: H pair}
H_\text{pair} = \sum_{\vect{k},\vect{k}'} V(\vect{k}-\vect{k}') \eta(\vect{k}) \sigma_1\eta(\vect{k}'),
\end{equation}
which is effective within the magnon Debye energy shell around the Fermi surface of the itinerant electron. The matrix $\sigma_1$ mix the pairing between the two pockets, or explicitly as $H_\text{pair}\sim V\eta_\Gamma\eta_M$. Because the pairing potential $V$ is positive, to gain the pairing energy, $\eta_\Gamma$ and $\eta_M$ are required
to take the opposite sign, which naturally leads to the $s_\pm$-wave pairing in the iron-based superconductor.

\begin{figure}[htbp]
\begin{center}
\includegraphics[width=0.32\textheight]{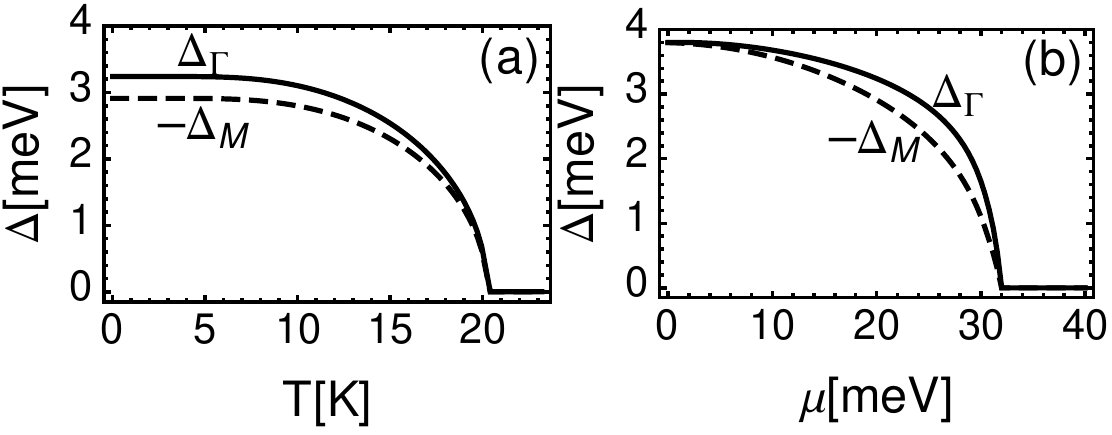}
\caption{(a) SC gaps v.s. temperature $T$, at $\mu=20$meV. (b) SC gaps v.s. chemical potential $\mu$, at $T=2$K. In our model, $\mu=30$meV is the level at which one pocket vanishes. The SC is suppressed due to the lack of the joint Fermi surface DOS. The SDW instability is not considered in this calculation.}
\label{fig: SC}
\end{center}
\end{figure}

The BCS theory for the multi-band SC (like in the iron-based superconductor) has been discussed in details in the literature,\cite{Dolgov09, Mazin09} but we will still briefly outline the main results here for the completeness of the present story. Let $\Delta_{\Gamma(M)}$ be the gap function around the $\Gamma(M)$ pocket, which could be arranged into a vector $\Delta=(\Delta_\Gamma,\Delta_M)^\intercal$. Following the standard treatment of the BCS theory,\cite{Manhan} the gap equation is given by
\begin{equation}
\Delta(\vect{k})= \sum_{\vect{k}'}V(\vect{k}-\vect{k}') \left(\begin{matrix}0 & f_M(\vect{k}') \\ f_\Gamma(\vect{k}') & 0\end{matrix}\right)\Delta(\vect{k}').
\end{equation}
where the pairing instability function reads
\begin{equation}
f_{\Gamma(M)}(\vect{k})=-\frac{1}{2E_{\Gamma(M)}(\vect{k})}\tanh\frac{\beta E_{\Gamma(M)}(\vect{k})}{2}.
\end{equation}
The gap function $\Delta$ can be determined self-consistently from the gap equation, which is found to be uniform in the momentum space with just a sign different between the $\Gamma$ and $M$ pockets. Their temperature and doping dependences are shown in \figref{fig: SC}. $\Delta_{\Gamma}$ and $-\Delta_M$ are close to each other in most cases, showing that the $s_\pm$-wave pairing is always dominant in the iron-based superconductor.

As the momentum dependence of $\Delta(\vect{k})$ is weak, the pairing potential can be approximated by its Fermi surface average $V_\text{SC}=\langle V(\vect{k}-\vect{k}')\rangle_{\vect{k}\vect{k}'\in\text{FS}}$ (which is positive). Close to the SC transition, the gap equation can be linearized to
\begin{equation}
\left(\begin{matrix}\Delta_\Gamma \\ \Delta_M\end{matrix}\right)
= -V_\text{SC}\ln\frac{\omega_D}{k_B T} \left(\begin{matrix}0 & N_M\\ N_\Gamma& 0 \end{matrix}\right)
\left(\begin{matrix}\Delta_\Gamma \\ \Delta_M\end{matrix}\right),
\end{equation}
where $N_{\Gamma(M)}$ stands for the Fermi surface density of state (DOS) on the $\Gamma(M)$ pocket. Here we have used $\sum_{\vect{k}}f_{\Gamma(M)}(\vect{k})\simeq -N_{\Gamma(M)}\ln\beta\omega_D$, as the pairing interaction is limited to a shell of $\omega_D$ around the Fermi surface. By diagonalization, the greatest eigen value is  $V_\text{SC}(N_\Gamma N_M)^{1/2}\ln\beta\omega_D$, which corresponds to the eigen vector $(\Delta_\Gamma, \Delta_M)\propto(N_\Gamma^{-1/2}, - N_M^{-1/2})$. The same result has been obtained in Ref.\,\onlinecite{Dolgov09}. The SC transition temperature can be estimated as
\begin{equation}
T_\text{SC}=1.136 \omega_D e^{-1/\lambda}
\end{equation}
The magnon Debye frequency $\omega_D\sim$50meV is higher than the phonon Debye frequency by one order of magnitude, which is why the iron-based superconductors can support a transition temperature beyond the McMillan limit. The pairing strength $\lambda = V_\text{SC} (N_\Gamma N_M)^{1/2}$ involves the Fermi surface DOS from both pockets. So the $s_\pm$-SC is suppressed at the band edge when either pocket vanishes under large doping (c.f. \figref{fig: SC}(b)).

\subsection{INS Spectrum}
In this section, we will discuss the inelastic neutron scattering spectrum of iron-based compounds. To simplify the notation, let us arrange the magnon field $n_b$, $n_c$ into the field $n=(n_b,n_c)^\intercal$, and define the propagator $G_n(q)=-\langle n(-q)n(q)\rangle$, which represents the generic spin-spin correlation in the system. Integrating out both the electron and spinon fields, we arrived at the dressed propagator of the spin fluctuation, according to the Dynson equation
\begin{equation}
G_n(q)^{-1}=
-\left(
\begin{matrix}
J_0^2\chi_c(q) & J_0\\
J_0 & J_0^2\chi_b(q)
\end{matrix}
\right),
\end{equation}
where the off diagonal term comes from the bare action $S_n\simeq J_0 n_b n_c$ in \eqnref{eq: bosonic model}, and $J_0^2\chi_{c,b}$ are self-energy corrections. On the one-loop level,
\begin{equation}\label{eq: chic chib}
\begin{split}
\chi_c(q) &= \Tr G_{c(0)}(k+q)s_c^\dagger G_{c(0)}(k)s_c,\\
\chi_b(q) &= -\Tr G_{b(0)}(k+q)s_b^\dagger G_{b(0)}(k)s_b,
\end{split}
\end{equation}
where $\chi_c$ ($\chi_b$) is the itinerant electron (spinon) spin susceptibility, reflecting the magnetic fluctuations in the Fermi-liquid (spin-liquid).

\begin{figure}[htbp]
\begin{center}
\includegraphics[width=0.28\textheight]{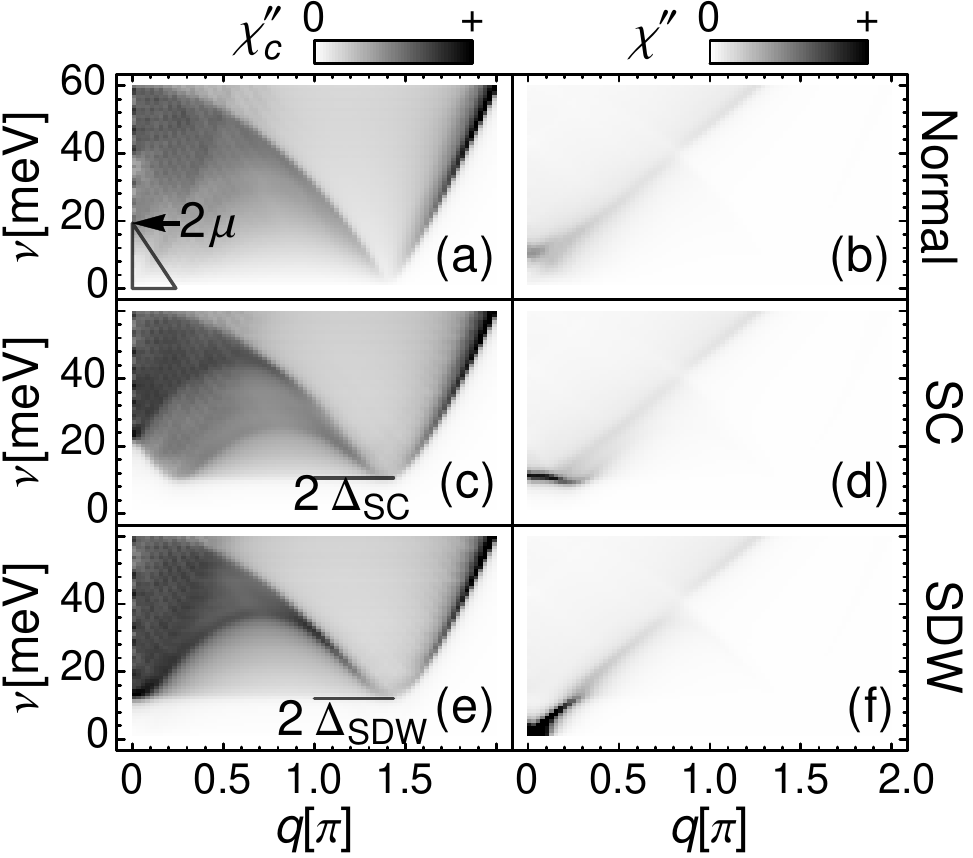}
\caption{The left column (a,c,e) shows the Stoner continuum of itinerant electron. The right column (b,d,f) shows the corresponding INS spectrum, calculated for $J_0=2.5$meV. The momentum origin is shifted to $M(\pi,0)$, i.e. $\vect{q}$ denotes the momentum divination from the $M$ point. (a,b) for normal state at 50K with $\mu=10$meV, (c,d) for $s_\pm$-SC state at 10K with $\mu=10$meV, $\Delta_\text{SC}=5$meV, and (g,h) for SDW state at 10K with $\mu=0$meV, $\Delta_\text{SDW}=6$meV. The spin resonance emerges at around 10meV in the SC phase (d), and is washed out by temperature as in (b). Gapless Goldstone mode appears in the SDW gap and continues to higher energy as in (f).}
\label{fig: INS}
\end{center}
\end{figure}

Because the neutron spin couples to the spin of both the itinerant electron $n_c$ and the local moment $n_b$, so the INS should probe the spectrum of both spin correlations $\chi(q)=\Tr G_n(q)$, which contains the contribution from both $\chi_c$ and $\chi_b$,
\begin{equation}\label{eq: chi}
\chi(q)=\frac{\chi_c(q)+\chi_b(q)}{1-J_0^2\chi_c(q)\chi_b(q)}.
\end{equation}
In the decoupled limit ($J_0\to 0$), \eqnref{eq: chi} reduces to $\chi(q) = \chi_c(q)+\chi_b(q)$, so the INS spectrum is a direct superposition of $\chi_c''(\nu,\vect{q})\equiv-2\Im\chi_c(q)|_{i\nu\to\nu+i0_+}$ and $\chi_b''(\nu,\vect{q})\equiv-2\Im\chi_b(q)|_{i\nu\to\nu+i0_+}$, i.e. $\chi''=\chi_c''+\chi_b''$. The contribution from the itinerant electron $\chi_c''$ has a dome-shaped Stoner continuum, as shown in \figref{fig: INS}(a), while the contribution from the spinon $\chi_b''$ contains quasi-sharp magnon modes, and was shown in \figref{fig: magnon}(a). It is found that $\chi_b''$ always overwhelms $\chi_c''$ by orders of magnitude, so in the decoupled limit ($J_0\to0$), the INS spectrum will just display the local moment fluctuation (the magnon spectrum). However if we turn on the coupling $J_0$, then the spin fluctuation of the itinerant electron will reshape the low energy part of the INS spectrum via the RPA correction in \eqnref{eq: chi}.

With finite coupling $J_0$, the spin fluctuation $\chi$ will be enhanced as $J_0$ reduces the denominator $(1-4J_0^2\chi_c\chi_b)$. New poles of $\chi$ could emerge when the denominator vanishes, leading to the spin resonance in the INS spectrum. There are mainly two factors that could induce the spin resonance around $\vect{Q}_s=(\pi,0)$ in the iron-based superconductors: (i) the bad nesting that clears up the Stoner continuum in a triangle region below $2\mu$ as in \figref{fig: INS}(a), (ii) the $s_\pm$-wave pairing that further gaps out the spectrum below $2\Delta$ as in \figref{fig: INS}(c). Both lead to the discontinuous jump of $\chi_c''$ at the lower edge of the continuum. According to the Kramers-Kronig relation, the real part of $\chi_c$ will get enhanced along this lower edge. So the resonance mode will first emerge right below the continuum, as shown in \figref{fig: INS}(d). The pairing driven resonance and its sensitivity on the pairing symmetry has been elaborated in the literature,\cite{Scalapino08a, Scalapino08b, Korshunov08, Bernevig09} so we will not repeat here. 

Above the SC transition temperature, the residual resonance signal\cite{Christianson08, Lumsden09, Qui09, Taylor11} can persist into the normal phase, but its intensity will be weakened by the thermal broadening effect, as in \figref{fig: INS}(b). In Ref.\,\onlinecite{YouPRB11}, it was proposed that the residual resonance is driven by the magnetic fluctuation between the mis-nested Fermi pockets, so its resonance energy is only controlled by doping, not affected by the SC gap, which explains its presence in the normal phase. In the two-fluid model, the low-energy spin resonance of the itinerant electron is usually carried on to the high-energy spin-wave mode of the local moment, thus it is possible to observe the spin resonance coexisting with the spin-wave in the INS spectrum.\cite{Qui09, Christianson09}

Towards the under-doped regime, both $\mu$ and $\Delta$ get smaller, and the resonance mode will get pushed to lower energy. We found that it will first be softened at finite momentum $\vect{q}$ deviated from $\vect{Q}_s$, leading to the incommensurate SDW\cite{Chubukov10} phase proximate to the SC phase on the under-doped side. The softened resonance mode will eventually evolves into the Goldstone mode in the SDW phase, as shown in \figref{fig: INS}(f). The gapless spin fluctuation reflects the spontaneous broken spin-rotational symmetry in the SDW phase, in agreement with the Goldstone's theorem.

Nevertheless in the joint ordering scenario of the two-fluid model, the microscopic origin of the gapless mode is non-trivial. The itinerant electron Fermi-liquid and the local moment spin-liquid are both \emph{gapped} in the SDW phase (see \figref{fig: SDW}(b) for spinon gap), but their collective spin dynamics is \emph{gapless}. The Goldstone mode is emergent on the RPA level. This was first pointed out in Ref.\,\onlinecite{YKW10} on a confined paramagnetic local moment background, using the Holstein-Primakoff boson language. Here we will briefly show that the emergent Goldstone mode still holds for the deconfined spin-liquid of local moment background, in the Schwinger boson language.

First of all, the SDW order parameters $n_b, n_c$ follow from the self-consistent equation \eqnref{eq: SDW Dynson},
\begin{equation}
\begin{split}
G_c(k)^{-1} &= G_{c(0)}(k)^{-1} + J_0 s_c n_b,\\
G_b(k)^{-1} &= G_{b(0)}(k)^{-1} + J_0 s_b n_c,
\end{split}
\end{equation}
from which $\partial_{n_b} G_c(k)= -G_c(k) J_0 s_c G_c(k)$, $\partial_{n_c} G_b(k) = -G_b(k) J_0 s_b G_b(k)$. While by definition, $n_b=-\sum_k\Tr G_b(k)s_b$ and $n_c=\sum_k\Tr G_c(k)s_c$. Take the derivative on both sides,
\begin{equation}\label{eq: partial n}
\begin{split}
\partial_{n_c}n_b&=J_0 \sum_k\Tr G_b(k) s_b G_b(k)s_b =-J_0\chi_b(0),\\
\partial_{n_b}n_c&=-J_0 \sum_k\Tr G_c(k) s_c G_c(k)s_c =-J_0\chi_c(0).
\end{split}
\end{equation}
Here we have used the definition of $\chi_b(q)$ and $\chi_c(q)$ in \eqnref{eq: chic chib}. $\partial_{n_c}n_b$ represent the response of $n_b$ to $n_c$, while $\partial_{n_b}n_c$ represents the response of $n_c$ to $n_b$. For a self-consistent mean-field (at the fixed point), we must have $(\partial_{n_c}n_b)(\partial_{n_b}n_c)=1$, i.e. the self-response of the order parameter is 1. According to \eqnref{eq: partial n}, such self-consistence condition is just equivalent to $1-J_0^2\chi_b(0)\chi_c(0)=0$. That is to say, if the SDW mean-field is self-consistent, $\chi(q)$ given in \eqnref{eq: chi} must diverge at $q=0$, which means the existence of the zero mode, and hence the spin fluctuation spectrum is gapless.

\begin{figure}[htbp]
\begin{center}
\includegraphics[width=0.27\textheight]{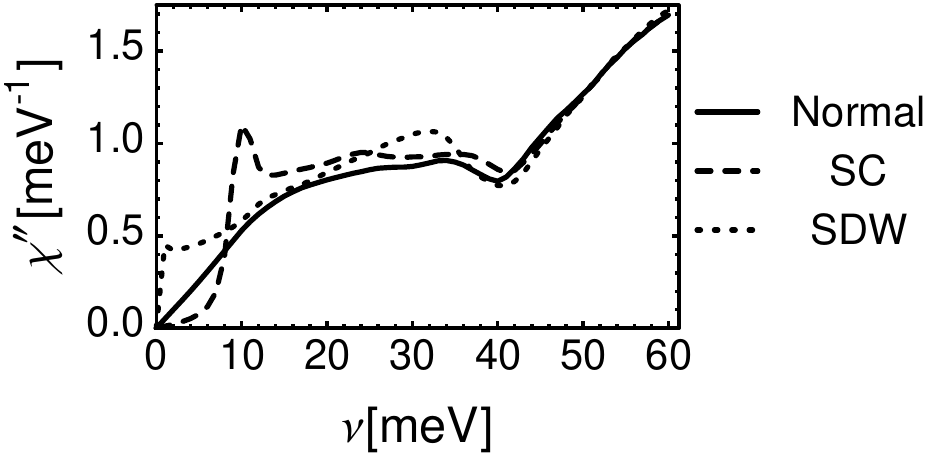}
\caption{The local spin fluctuation $\chi''(\nu) = \sum_{\vect{q}}\chi''(\nu,\vect{q})$, in various phases: the normal phase (solid curve) corresponding to \figref{fig: INS}(b), the SC phase (dashed curve) corresponding to \figref{fig: INS}(d), the SDW phase (dotted curve) corresponding to \figref{fig: INS}(f). The peak at around 10meV manifests the spin-resonance mode in the SC phase. The kink around 40meV indicates the upper edge of the Stoner continuum, above which the spectrum is not affected by the itinerant electron.}
\label{fig: stoner}
\end{center}
\end{figure}

The local spin fluctuation spectrum $\chi''(\nu)=\sum_\vect{q}\chi''(\nu,\vect{q})$ is readily obtained from the INS spectrum $\chi''(\nu,\vect{q})$ via the momentum summation. The results for various phases are shown in \figref{fig: stoner}. Compare to the normal phase, the low-energy spin fluctuation is suppressed in the SC phase, and is enhanced in the SDW phase. The spin-resonance in the SC phase and the gapless spectrum in the SDW phase are clearly demonstrated here. 

However a universal kink structure at about 40meV in \figref{fig: stoner} worth further discussion. This energy scale comes from the upper boundary of the Stoner continuum of the itinerant electron as in \figref{fig: INS}(a). This singularity is reflected as a little kink in all spectrums on the right column of \figref{fig: INS}. This kink separates the spin fluctuation spectrum into the high and low energy parts. According to our two-fluid model, the low energy part is affected by the itinerant electron, while the high energy part reflects the underlying local moment fluctuation. In different phases, the local moment fluctuation does not change much, but the itinerant electron part can be quite different, so that the changes of the spectrum mainly happen in the low energy regime, as demonstrated in \figref{fig: stoner}, unlike the conventional AFM material, where the low energy fluctuation is originated from the local moment and the Stoner continuum appears at high energy. This peculiar spin fluctuation spectrum predicted\cite{YouPRB11} by the two-fluid model has been observed in the INS experiment.\cite{PDai12}

\section{Charge Dyanmics}
\subsection{Fermionic Decomposition}
Now we turn to the other decomposition channel of the electron-spinon interaction: the fermionic channel, and study the charge dynamics of the two-fluid model.

Starting from the interaction Hamiltonian $H_{cb}$ in \eqnref{eq: Hcb}, decompose it by introducing the fermionic field $f_{\eta\sigma_f}$. Here $\eta=p,e$ is the channel index, labeling the two Fermionic channels: the pairing channel $p$ and the exchange channel $e$. In the exchange channel, the itinerant electron emits a spinon and becomes the composite fermion $c_{\sigma_c}b^\dagger_{\bar{\sigma}_b}\to f_{e\sigma_f}$. In the pairing channel, one itinerant electron merges with one spinon to form composite fermion $c_{\sigma_c}b_{\sigma_b}\to f_{p\sigma_f}$. The spin of the composite fermion is labeled by the combined index $\sigma_f = \sigma_c\sigma_b$ which has 4 possibilities $\sigma_f = \uparrow\uparrow, \uparrow\downarrow, \downarrow\uparrow, \downarrow\downarrow$. As a combination of two spin-1/2 objects, the composite fermion has an integer spin, including a singlet state (spin-0) and three spin triplet states (spin-1).

Through a fermionic version of the Hubbard-Stratonovich (HS) transform, $H_{cb} \to H_{fcb}+H_{f}$
\begin{equation}\label{eq: Hfcb}
\begin{split}
H_{fcb} = \frac{J_0}{2} &\sum_{\vect{K}_b,\vect{k}_b,\sigma_b}\sum_{\vect{K}_c,\vect{k}_c,\sigma_c}\\
& (-)^{\sigma_b} f^\dagger_{e(\vect{K}_f+\vect{k}_f)\sigma_f}c_{(\vect{K}_c+\vect{k}_c) \sigma_c}b^\dagger_{-(\vect{K}_b+\vect{k}_b) \bar{\sigma}_b}\\
& + f^\dagger_{p(\vect{K}_f+\vect{k}_f)\sigma_f}c_{(\vect{K}_c+\vect{k}_c) \sigma_c}b_{(\vect{K}_b+\vect{k}_b) \sigma_b}+h.c.,
\end{split}
\end{equation}
where $\vect{K}_f=\vect{K}_b+\vect{K}_c$, $\vect{k}_f=\vect{k}_b+\vect{k}_c$ follows from the momentum conservation law.
\begin{equation}
\begin{split}
H_f=\frac{J_0}{2} \sum_{\vect{K}_f,\vect{k}_f}& f_{p(\vect{K}_f+\vect{k}_f)\sigma_f}^\dagger g_{\sigma_f\sigma'_f}f_{p(\vect{K}_f+\vect{k}_f)\sigma'_f}\\
& - f_{e(\vect{K}_f+\vect{k}_f)\sigma_f}^\dagger g_{\sigma_f\sigma'_f} f_{e(\vect{K}_f+\vect{k}_f)\sigma'_f}.
\end{split}
\end{equation}
Here $g$ is a $4\times4$ matrix defined as the inverse of kernel matrix of Hund's rule coupling $g \equiv (\sum_{i=1,2,3}\sigma_{ii})^{-1}$. Integrating out the composite fermion $f_{\eta\sigma_f}$ degrees of freedom would restore the Hund's rule interaction Hamiltonian $H_{cb}$.

For now, $f_{\eta\sigma_f}$ is just an auxiliary field in the HS transform. Its spectrum is featureless without quasiparticle peak. However as we integrate out the electron and the spinon fields, the composite fermion $f$ will acquire its dynamics from the collective motion of the electron and the spinon, and the quasiparticle can emerge in the spectrum of $f$ as the new poles via the RPA approach. In that case, with an intermediate coupling strength of the effective Hund's rule interaction, the auxiliary field $f$ will become a well-defined low energy quasiparticle composed of an electron and a spinon.

\subsection{Emergent Composite Fermion}

To demonstrate the emergence principle of the composite fermion, let us start with a momentum-independent toy model, neglecting all the pockets and spin degrees of freedom. The toy model is described by the Hamiltonian $H=H_{c}+H_{b}+H_{cb}$, in which a single-level electron couples to a single-mode spinon by an attractive interaction $-J_0$,
\begin{equation}\label{eq: toy model}
\begin{split}
H_{c}&=c^\dagger \epsilon c,\\
H_{b}&=-\frac{1}{2}\eta(bb+h.c.)+\lambda b^\dagger b,\\
H_{cb}&=-J_0 c^\dagger c b^\dagger b.
\end{split}
\end{equation}
By HS decomposition in the fermionic channels, $H_{cb}$ can be cast into $H_{f}+H_{fbc}$,
\begin{equation}
\begin{split}
H_{f}&=\frac{J_0}{2}(f_p^\dagger f_p+f_e^\dagger f_e),\\
H_{fcb}&=\frac{J_0}{2}(f_p^\dagger cb + f_e^\dagger c b^\dagger +h.c.),
\end{split}
\end{equation}
where two flavors of composite fermions $f_{p(e)}$ are introduced for the pairing (exchange) channel.

Arrange the field variables into $\psi_c=c$, $\psi_b = (b,b^\dagger)^\intercal$ and $\psi_f=(f_p,f_e)^\intercal$, and switch to the path integral formalism, the partition function reads $Z=\int\mathcal{D}[\psi_c,\psi_b,\psi_f] e^{-S}$, with the action $S=S_c+S_b+S_f+S_{fcb}$ given by
\begin{equation}
\begin{split}
S_c&=\sum_{i\omega_c}\psi_c^\dagger(-i\omega_c+\epsilon)\psi_c,\\
S_b&=\frac{1}{2}\sum_{i\omega_b}\psi_b^\dagger(-i\omega_b\sigma_3+\lambda\sigma_0-\eta\sigma_1)\psi_b,\\
S_f&=\frac{J_0}{2}\sum_{i\omega_f}\psi_f^\dagger \sigma_0\psi_f,\\
S_{fcb}&=\frac{J_0}{2}\sum_{i\omega_f,i\omega_c,i\omega_b}(\psi_f^\dagger\sigma_0\psi_c\otimes\psi_b + h.c.)\\
&\hspace{8em}\delta(i\omega_f=i\omega_c+i\omega_b).
\end{split}
\end{equation}
Here $i\omega_{c/b/f}$ denotes the Matsubara frequency for the field $\psi_{c/b/f}$. The propagators are defined as $G_{c/b/f}=-\langle\psi_{c/b/f}\psi_{c/b/f}^\dagger\rangle$, represented by the solid/dotted/dashed line, whose bare forms are readily obtained by inverting the action kernel
\begin{equation}
\begin{split}
G_{c(0)}(i\omega_c)&=-\diag{fig_Gc0}{8pt}=\frac{1}{i\omega_c-\epsilon},\\
G_{b(0)}(i\omega_b)&=-\diag{fig_Gb0}{8pt}=2\frac{i\omega_b\sigma_3+\lambda\sigma_0+\eta\sigma_1}{(i\omega_b)^2-\lambda^2+\eta^2},\\
G_{f(0)}&=-\diag{fig_Gf0}{8pt}=-2\sigma_0/J_0.
\end{split}
\end{equation}
The bare propagator $G_{f(0)}$ for the composite fermion is featureless. Integrate out the $\psi_c$ and $\psi_b$ field brings correction to it. On the one loop level, the Dyson equation reads $\diag{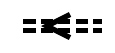}{8pt}=\diag{fig_Gf0}{8pt}+\diag{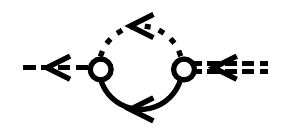}{20pt}$, whose solution can be formally written as
\begin{equation}\label{eq: Gf toy formula}
\begin{split}
G_f&=-\diag{fig_Gf}{8pt}=-\big(\diag{fig_Gf0}{8pt}^{-1}-\diag{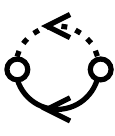}{20pt}\big)^{-1}\\
&=(G_{f(0)}^{-1}-\Sigma_f)^{-1},
\end{split}
\end{equation}
with the self-energy (in the low temperature $T\to0$ limit)
\begin{equation}
\begin{split}
\Sigma_f(i\omega_f)&=-\diag{fig_Sf}{20pt}\\
&=-\frac{J_0^2}{4}\sum_{i\omega_b} G_{c(0)}(i\omega_f-i\omega_b)\otimes G_{b(0)}(i\omega_b)\\
&=\frac{J_0^2}{4}\frac{(\lambda/\Omega)\sigma_0+(\eta/\Omega)\sigma_1+s_\epsilon\sigma_3}{i\omega_f-s_\epsilon(|\epsilon|+\Omega)},
\end{split}
\end{equation}
where $\Omega=\sqrt{\lambda^2-\eta^2}$ and $s_\epsilon\equiv\sgn\epsilon=\pm1$ denotes the sign of $\epsilon$. Substitute it into \eqnref{eq: Gf toy formula}, the composite fermion propagator is dressed to
\begin{equation}\label{eq: Gf toy solution}
G_f(i\omega_f)=-\frac{2\sigma_0}{J_0}+\frac{(\lambda/\Omega)\sigma_0+(\eta/\Omega)\sigma_1+s_\epsilon\sigma_3}{i\omega_f-s_\epsilon(|\epsilon|+\Omega)+J_0(\lambda/\Omega)},
\end{equation}
with new poles appearing at the frequency $i\omega_f=s_\epsilon(|\epsilon|+\Omega)-J_0(\lambda/\Omega)$ due to the self-energy correction.

\begin{figure}[htbp]
\begin{center}
\includegraphics[width=0.24\textheight]{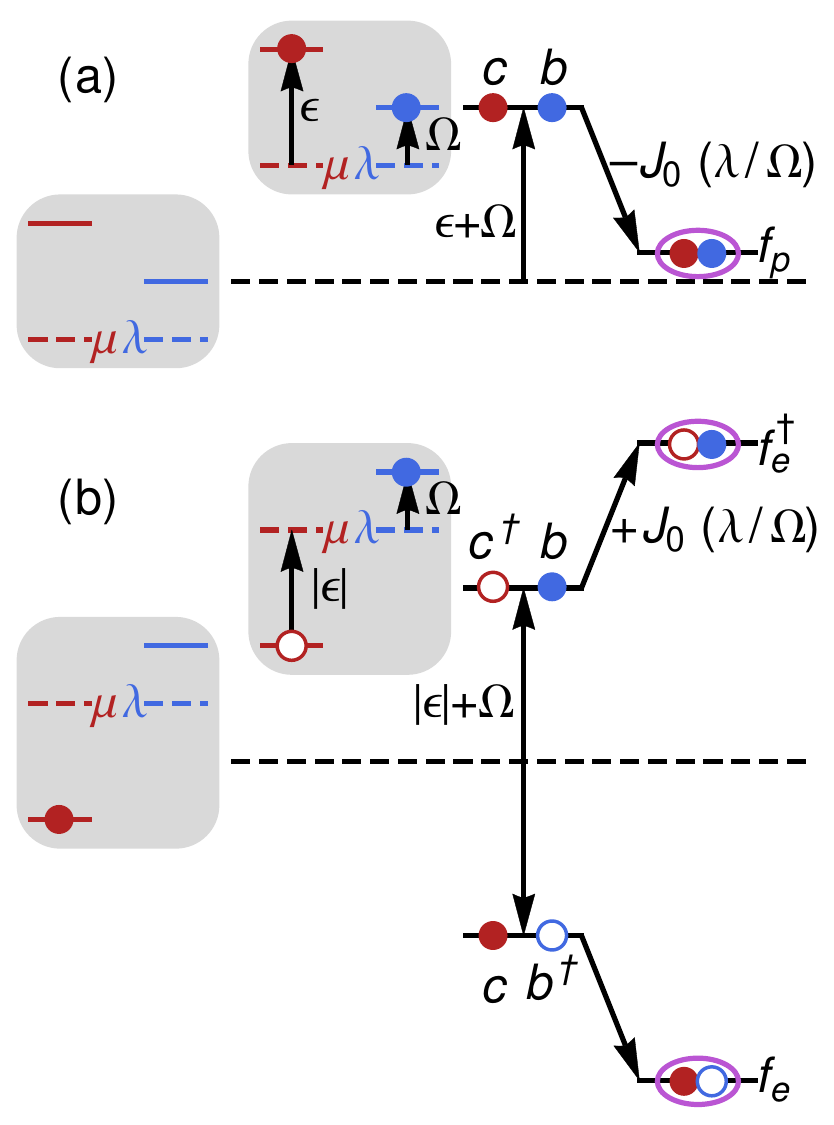}
\caption{The formation of the composite fermion. (a) the $\epsilon>0$ case, (b) the $\epsilon<0$ case. The black dashed line indicates the ground state energy of the electron-spinon many-body system. The gray-background insets depict the energy-level-filling of the electron (in red) and the spinon (in blue). $\mu$ ($\lambda$) marks the chemical potential level of the electron (spinon). In the virtual process (a), a pair of electron and spinon is first excited by $\epsilon+\Omega$, and then they attract to form the composite fermion $f_p$ releasing the binding energy $-J_0(\lambda/\Omega)$. In the virtual process (b), first excite a pair of hole and spinon from the ground state by $|\epsilon|+\Omega$, and further absorb $J_0(\lambda/\Omega)$ energy to overcome the repulsion between them to form the composite fermion $f_e^\dagger$, whose particle-hole conjugation $f_e$ is of the opposite energy.}
\label{fig: composite fermion}
\end{center}
\end{figure}

According to the Hamiltonian $H_{cb}$ in \eqnref{eq: toy model}, the electron $c$ and spinon $b$ attract each other. When the electron energy level is above the Fermi energy $\epsilon>0$ ($s_\epsilon=+1$), new pole emerge in the $f_p$ channel at the energy $E_{p}=\epsilon+\Omega-J_0(\lambda/\Omega)$ as illustrated in \figref{fig: composite fermion}(a). When the electron level is below the Fermi energy $\epsilon<0$ ($s_\epsilon=-1$), new pole emerge in the $f_e$ channel at the energy $E_{e}=-|\epsilon|-\Omega-J_0(\lambda/\Omega)$ as illustrated in \figref{fig: composite fermion}(b). 

\begin{figure}[htbp]
\begin{center}
\includegraphics[width=0.28\textheight]{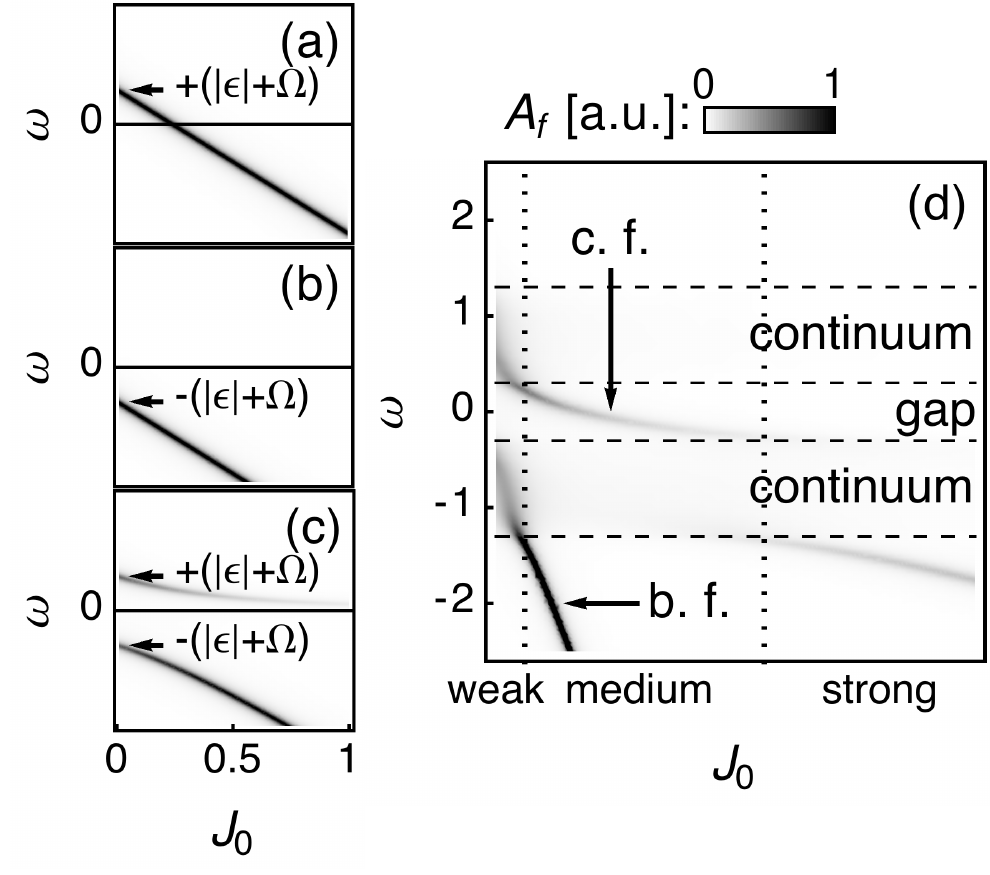}
\caption{Spectral function of the composite fermions under different coupling strength $J_0$. (a) The $\epsilon>0$ case: the $f_p$ mode energy $E_p$ decreases with $J_0$. (b) The $\epsilon<0$ case: the $f_e$ mode energy $E_e$ decreases with $J_0$. (c) With both electron levels $\pm\epsilon$ presented: the two branches are bent due to the level repulsion. (d) Gapped spinon coupled to a band of electrons. The horizontal dashed line marks out the edge of the electron-spinon joint continuum. The vertical dotted line separates the weak, medium and strong coupling regions. With increasing coupling $J_0$, the tightly-bound fermion (b.f.) is quickly pushed to high energy, while the composite fermion (c.f.) emerges in the spinon gap and is weaken for larger $J_0$.}
\label{fig: toy}
\end{center}
\end{figure}

If we include two electron levels both above and below the Fermi energy,  then both the $f_p$ and the $f_e$ modes will emerge, and one may expect their energies to decrease linearly with $J_0$ as $E_{p(e)}=\pm(|\epsilon|+\Omega)-J_0(\lambda/\Omega)$, as shown in \figref{fig: toy}(a, b). However there is a mixing between the $f_p$ and the $f_e$ modes, which is induced by the pairing of the spinon (as the pairing mixes $b\leftrightarrow b^\dagger$ and hence the composite fermions $f_p\simeq c b\leftrightarrow f_e\simeq cb^\dagger$). The mixing of $f_p$ and $f_e$ leads to the level repulsion between $E_p$ and $E_e$, such that the $E_p$ branch will be bent, see \figref{fig: toy}(c), and stop decreasing around the Fermi energy, resulting in a low energy mode.

If we further include a band of electron levels across the Fermi energy, of the band width $W$, we will see that the electron-spinon joint continuum in the composite fermion spectrum is separated by the spinon gap. In the weak coupling limit $J_0 \simeq 0$, the spectral weight is transferred within the joint continuum, see \figref{fig: toy}(d). However when the coupling strength grows beyond a certain value $J_0 \gtrsim W\Omega/\lambda $, two modes emerge in the spectrum as in \figref{fig: toy}(d): (i) a tightly-bound fermion mode will appear outside the continuum, of the spectral weight $\sim 1$, and being pushed all the way up to the high energy sector with increasing $J_0$; (ii) a composite fermion mode will emerge within the spinon gap of fractional spectral weight $\sim J_0^{-2}$ which is gradually weaken and transferred to the high energy sector. The tightly-bound fermion corresponds to a state that the electron is strongly locked to the spinon and becomes a part of the local moment, which represents the renormalization of the local moment by the itinerant electron. Moreover, a fraction of the electron spectral weight is leftover within the spinon gap, resulting in the in-gap composite fermion mode. Upon further increase of $J_0$, the composite fermion mode will eventually fade away, and the electron will be completely locked with the spinon. So we conclude from the above analysis of the toy model that if we couple a band of fermions to some gapped bosons, there exists an \emph{intermediate} coupling region, in which the low-energy resonance mode will appear as the \emph{composite fermion} within the boson gap.

\subsection{Composite Mode and STS Spectrum}

Now we come back to the momentum-dependent version of the two-fluid model, and put in the pocket (valley) and spin degrees of freedom. We may switch to the field theoretical notation. Introduce the composite fermion field $\psi_f$ by arranging the $f$ fermions following the order of particle-hole, pairing-exchange, valley, and spin degrees of freedom,
\begin{equation}\label{eq: psi_f}
\psi _f=\left(
\begin{smallmatrix}
 f_k \\
 \mathcal{T} f_{-k}^{\dagger }
\end{smallmatrix}
\right)\otimes \left(
\begin{smallmatrix}
 f_p \\
 f_e
\end{smallmatrix}
\right)\otimes \left(
\begin{smallmatrix}
 X \\
 Y
\end{smallmatrix}
\right)\otimes \left(
\begin{smallmatrix}
 \uparrow  \\
 \downarrow 
\end{smallmatrix}
\right)_c\otimes \left(
\begin{smallmatrix}
 \uparrow  \\
 \downarrow 
\end{smallmatrix}
\right)_b.
\end{equation}
Then $e^{-S_{cb}}=\int\mathcal{D}[\psi_f]e^{-S_f-S_{fcb}}$,
\begin{equation}
\begin{split}
S_f &=-\sum _k \psi _f^{\dagger }(k)G_{f(0)}(k)^{-1}\psi _f(k)\\
S_{fcb} &=\sum _{k, k'} \Lambda  \psi _f\left(-k-k'\right)\psi _c(k)\psi _b\left(k'\right)+h.c.,
\end{split}
\end{equation}
where the bare propagator is
\begin{equation}
G_{f(0)}(k)=-4J_0^{-1}\sum_{i=1,2,3}\sigma_{330ii}.
\end{equation}
The vertex operator $\Lambda$ is a three-leg tensor, whose indices $\alpha_f$, $\alpha_c$ and $\alpha_b$ label the field components of $\psi_f$, $\psi_c$ and $\psi_b$ respectively. The tensor is written down according to the vertex Hamiltonian \eqnref{eq: Hfcb}, or explicitly reads
\begin{equation}
\Lambda_{\alpha_f\alpha_c\alpha_b}=\frac{J_0}{4}(-)^{h_c h_b+s_c+s_b},
\end{equation}
with $\alpha_f=16(1-h_c)+8(h_b+h_c)_{\mod2}+4(K_c+K_b)_{\mod 2}+2(1-s_c)+(1-s_b)+1$, $\alpha_c=4h_c+2K_c+s_c+1$,  $\alpha_b=4h_b+2K_b+s_b+1$, with $h_{c,b}, K_{c,b}, s_{c,b}$ enumerated over 0, 1. 

\begin{figure}[htbp]
\begin{center}
\includegraphics[width=0.24\textheight]{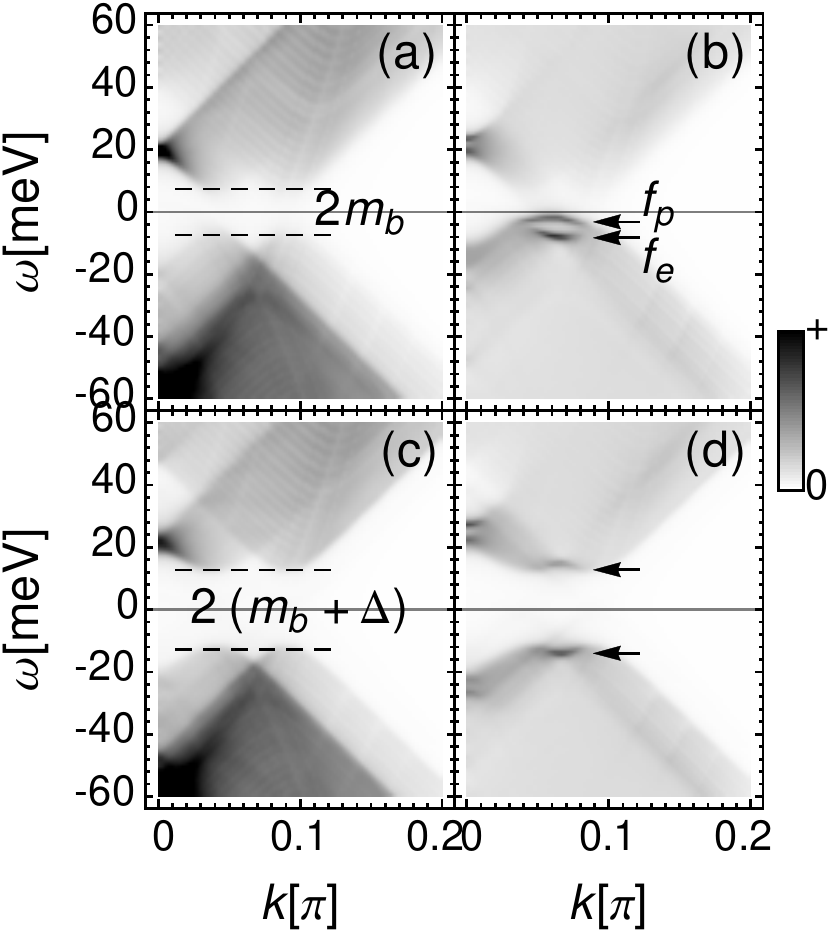}
\caption{Left column: the spectrum of the self-energy $-2\Im\Sigma_f$ in (a) normal phase and (c) SC phase. The dashed lines mark the edges of the electron-spinon continuum. Right column: the spectrum of the composite fermion $-2\Im G_f$ in (b) normal phase and (d) SC phase. The arrows point out the composite fermion modes. In all figures, the $(\sigma_{00000}+\sigma_{30000})$ component (the particle channel) is shown, under the condition $T=20$K and $\mu=20$meV. The SC gap is taken to be 5meV.}
\label{fig: SFDFmap}
\end{center}
\end{figure}

Following the same approach discussed in the previous subsection, we integrate out the electron and the spinon fields to study the effective theory for composite fermions. The self-energy correction for the composite fermion comes from the electron-spinon bubble diagram
\begin{equation}\label{eq: Sf}
\Sigma_f(k)=-\diag{fig_Sf}{20pt}=-\sum_q \Tr G_{c(0)}(k-q)\Lambda^\dagger G_{b(0)}(q)\Lambda.
\end{equation}
Its spectrum $-2\Im \Sigma_f(k)|_{i\omega\to\omega+i0_+}$ is shown in \figref{fig: SFDFmap}(a), which reflects the  density of states of the electron-spinon joint excitations. The electron-spinon continuum is gapped away from the Fermi level by $\pm m_b$. Because to excite the spinon in the spin-liquid requires at least $m_b$ amount of energy to overcome the spinon gap, while the itinerant electron is gapless, so their joint excitations only appear outside the energy range $\pm m_b$. Due to the jump of the spectrum at the edge of the electron-spinon continuum, the real part of the self-energy $\Sigma_f$ will be enhanced there according to the Kramers-Kronig relation, which will lead to the composite fermion resonance mode to appear via the RPA approach.

Following the similar Dynson's equation \eqnref{eq: Gf toy formula}, the bare propagator of $\psi_f$ field is dressed as
\begin{equation}\label{eq: Gf}
G_f(k)=(G_{f(0)}^{-1}(k)-\Sigma_f(k))^{-1}.
\end{equation}
From the dressed propagator, one may extract its spectrum $A_f(\omega,\vect{k})=-2\Im G_f(k)|_{i\omega\to\omega+i0_+}$, as shown in \figref{fig: SFDFmap}(b). With an intermediate coupling strength $J_0=50$meV, the in-gap composite mode emerges from the edge of the continuum, and is pushed towards the Fermi level for larger $J_0$. This is because $\Sigma_f$ is enhanced at the edge, so the denominator of \eqnref{eq: Gf} will first approach to zero there, which leads to new poles in the originally featureless propagator $G_f$.

As a bound state of the electron and the spinon, each composite fermion carries one electron charge, which should contribute to the charge transport and the electromagnetic response of the material, and could be probed in the scanning tunneling microscope (STM) experiment. 

\begin{figure}[htbp]
\begin{center}
\includegraphics[width=0.2\textheight]{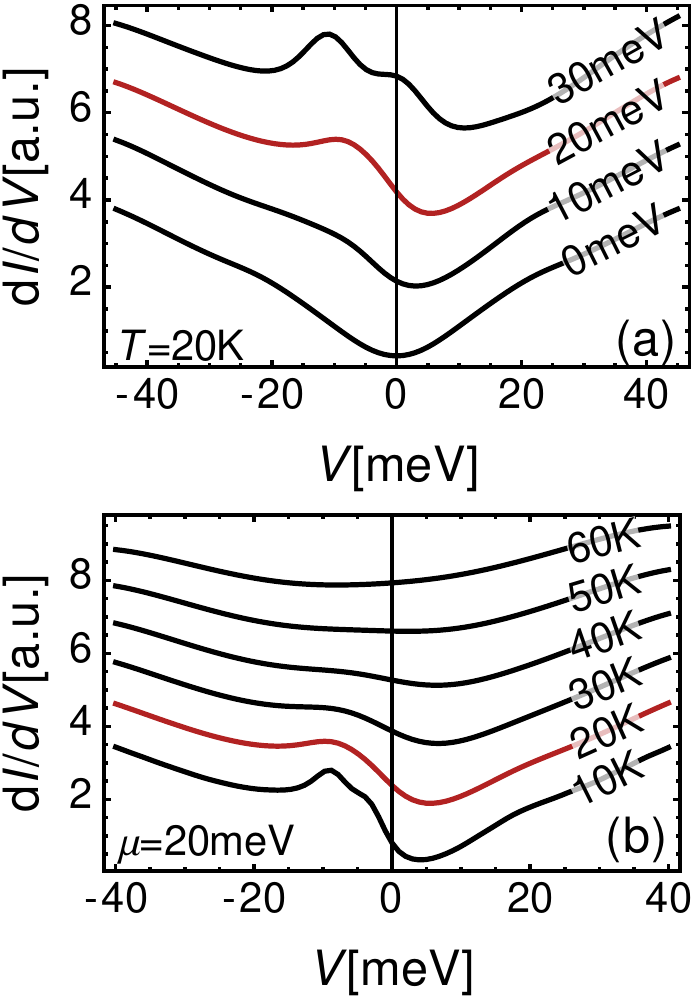}
\caption{The composite fermion mode contribution to the STS spectrum. (a) For different chemical potentials $\mu$ of the itinerant electron. (b) Under different temperatures $T$. Curves are offset vertically for clarity. The typical curve with $T=20$K and $\mu=20$meV is marked out in red.}
\label{fig: dIdV}
\end{center}
\end{figure}

Let us consider the STM differential conductance $\mathrm{d}I/\mathrm{d}V$, also known as the scanning tunneling spectroscopy (STS) measurement. When the electron is injected to the material from the STM tip, it can grasp a spinon to form the composite fermion. Given the inhomogeneity observed in many iron-based superconductors, we may assume local condensation of spinons, i.e. $\langle b\rangle \neq 0$ locally. If the tip happens to be suspended above the local condensate, the injected electron can immediately become composite fermion $f_p=\langle b\rangle c$ or $f_e=\langle b^\dagger\rangle c$, and the electric current is carried on by the composite fermion. Therefore the composite fermion should also contribute to the differential conductance $\mathrm{d}I/\mathrm{d}V$ spectrum,
\begin{equation}
\frac{\mathrm{d}I}{\mathrm{d}V}\propto\sum_{\omega,\vect{k}}n_F^\prime(V-\omega)A_f(\omega,\vect{k}),
\end{equation}
where $n_F^\prime(\omega)=(\beta/4)\text{sech}^2(\beta\omega/2)$ is the derivative of the Fermi distribution. A typical spectrum is shown in red in \figref{fig: dIdV}, which has an  asymmetric hump-dip feature with a hump below the Fermi level and a dip near the Fermi level, dubbed as the ``pseudo-gap'' in literature.\cite{YYWang12} The hump corresponds to the emergent composite fermion mode marked out by arrows in \figref{fig: SFDFmap}(b). Its energy scale is controlled by the spinon gap $m_b$ at around 10meV. While the dip is due to the lack of composite fermion spectrum in the gap of the electron-spinon continuum. So the hump-dip feature in the STS spectrum is nothing but the gap of composite fermions (electron-spinon joint excitations), which is naturally pinned at the Fermi level (as controlled by the spinon gap), and will not shift with the electron doping.

The asymmetric STS line shape about the Fermi level is due to the finite doping away from the perfect nesting level (modeled by the finite $\mu=20$meV here). From \figref{fig: dIdV}(a), one can see the gap feature starts out symmetric at perfect nesting ($\mu=0$meV), and becomes more and more asymmetric as $\mu$ shifting away. As the Fermi level deviates from the perfect nesting, the electron and hole bands are not symmetric to each other about the Fermi level, which leads to the asymmetric electron-spinon continuum and hence the asymmetric composite fermion spectrum. For electron doping, the electron-spinon joint spectral weight below the Fermi level is stronger as in \figref{fig: SFDFmap}(a), which will push the composite mode upwards under the level-repulsion effect. So the composite mode at the lower edge will be pushed into the gap and becomes a resonance mode, while that at the upper edge will be pushed into the continuum and damp out. Thus we conclude that the electron (hole) doping causes the composite fermion mode to emerge from below (above) the Fermi level (see \figref{fig: SFDFmap}(b) for the electron doped case). This doping dependence of the composite fermion energy has been observed recently. In the electron doped NaFe$_{0.94}$Co$_{0.06}$As sample,\cite{YYWang12} the hump structure (signature of the composite fermion) appears below the Fermi level; while in the hole doped Ba$_{0.6}$K$_{0.4}$Fe$_2$As$_2$ sample,\cite{Wang2013} the hump structure appears above the Fermi level. Comparing both experiments Refs.\,\onlinecite{YYWang12} and \onlinecite{Wang2013}, the low-energy STS spectrum (within the range of $\pm50$meV) is particle-hole reversed with respect to the Fermi level between the electron and hole doped cases, as expected in our theory.

From \figref{fig: dIdV}(a), we can see the composite fermion mode is becoming stronger for larger doping (as the mode is pushed deeper into the gap by stronger level-repulsion), and is expect to be more obvious in the over-doped region. Upon raising the temperature, the hump-dip feature will eventually be smeared out by the thermal broadening as in \figref{fig: dIdV}(b). These features are also consistent with the observation\cite{YYWang12} in the 111-type compounds. 

\subsection{Influence of SC and Magnetic Field}

As a resonance mode of the coupled itinerant electron and the spinon, the composite fermion will be influenced by both the components of the Fermi-liquid and the spin-liquid in the two-fluid description. In this section, we will discuss how the composite fermion responses to the itinerant electron pairing and the external magnetic field.

Consider the $s_\pm$-wave pairing of the itinerant electron, with SC gap $\Delta_\text{SC}$. Then the itinerant electron is no longer gapless excitations: at least an additional energy of the amount of $\Delta_\text{SC}$ must be paid to excite the electron. So the electron-spinon joint excitation energy is raised to above $(m_b+\Delta_\text{SC})$, i.e. the joint gap is enlarged to $\pm(m_b+\Delta_\text{SC})$, as in \figref{fig: SFDFmap}(c). So the itinerant electron pairing would enlarge the hump-dip feature in the STS spectrum.

To verify this, we take $\Delta=5$meV (pairing gap) of $s_\pm$ symmetry, and modify the itinerant electron propagator to
$G_{c(0)}(k) =2(i\omega\sigma_{000}+\mu\sigma_{300}+\epsilon(\vect{k})\sigma_{330}+\Delta_\text{SC} \sigma_{130})^{-1}$. Re-calculate the composite fermion self-energy $\Sigma_f$ according to \eqnref{eq: Sf}, whose spectral function $\Sigma_f(k)|_{i\omega\to \omega+i0_+}$ is shown in \figref{fig: SFDFmap}(c). Substitute $\Sigma_f$ into \eqnref{eq: Gf}, we can obtain the spectrum of the composite fermion in the SC phase, as \figref{fig: SFDFmap}(d). The composite modes appear at the edge of the joint gap, as indicated by the arrows.

\begin{figure}[htbp]
\begin{center}
\includegraphics[width=0.22\textheight]{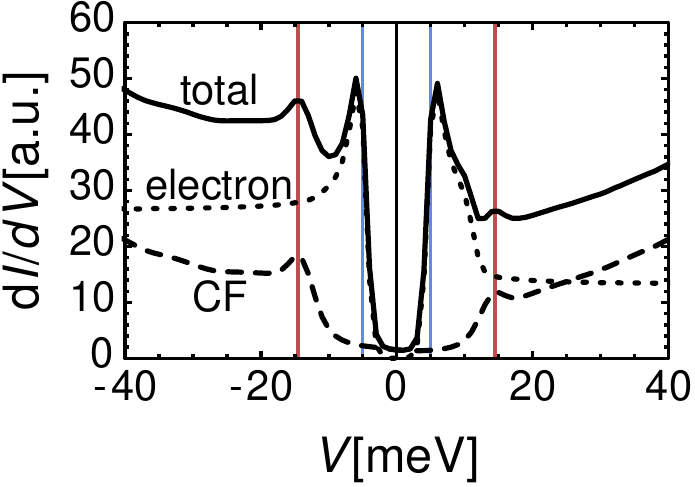}
\caption{The STS spectrum in the SC phase. The total differential conductance (solid curve) includes the contribution from the itinerant electron (dotted curve) and the composite fermion (CF, dashed curve). The blue vertical line marks the SC gap $\pm\Delta_\text{SC}$, and the red vertical line marks the position of the composite mode $\pm(m_b+\Delta_\text{SC})$. The calculation was done with $\mu=20$meV, $\Delta_\text{SC}=5$meV at $T=5$K.}
\label{fig: twogap}
\end{center}
\end{figure}

Because both the itinerant electron and the composite fermion mode can conduct electric current, the STM differential conductance is a sum of both contributions, as shown by the solid curve in \figref{fig: twogap}. Here the itinerant electron contribution is shown as the dotted curve, which exhibits the $\pm\Delta_\text{SC}$ symmetric pairing gap around the Fermi surface (a sudden drop at +10meV is due to the $\Gamma$ band top, which is not a universal feature). While the dashed curve is the contribution of the composite fermion, which displays the asymmetric hump-dip feature with resonance composite modes at the gap edges. Putting two contributions together, we arrive at the two-gap feature in the STS spectrum, which contains two extra little peaks outside the SC coherence peaks. The extra peaks from the composite fermion modes has been observed in the recent experiment.\cite{Wang2013}

Now we consider the response of the composite fermion mode to the external magnetic field. First of all, the composite fermion spin can be either in a triplet state or in a singlet state. For the spin-triplet composite fermion, its energy level will Zeeman split in the external magnetic field. For the spin-singlet composite fermion, its weight will be suppressed by the magnetic field, because the Zeeman splitting of the spinon spectrum will reduce and smear out the spinon gap, such that the electron-spinon continuum could expand toward the Fermi level and damp out the composite fermion. Therefore we predict that with the external magnetic field, the hump-dip structure in the STS spectrum will be weaken, and the extra peaks observed in the SC phase will split and/or weaken.

\subsection{Other Theoretical Possibilities}

Admittedly the two-fluid model we have presented here may not be the only theory to explain the observed hump-dip feature and extra peaks in the STS spectrum. There are still other theoretical possibilities, which we will briefly review/comment as follows.

(a) Band structure origin. This theory explains the STS hump structure by the itinerant electron band structure. Indeed, ARPES experiments have commonly observed a shallow flat band 10meV below the Fermi level around the $\Gamma$ point in many iron-based compounds,\cite{Liu10,Liu11,Richard11,Zhang12} which coincides with the hump structure in the energy scale. But there are several issues. First, if the hump-dip feature comes from the band structure, it must exhibit rigid band shift under doping, which contradicts with the observation. Second, the STS line shape is particle-hole reversed about the Fermi level in the complementary experiments of electron/hole doping, which is hard to explain by the electron band structure. Third, because electron band is a momentum space structure, so its signal should be rather uniform in the real space. However the observed hump-dip feature is strongly inhomogeneous, which is more likely to originate from some local degrees of freedom.

(b) SDW reentrance.\cite{Jiang12} This theory assumes a second SDW phase in the over-doped region, driven by the nesting of the electron pocket with the second hole pocket, such that the hump-dip feature could be understood as the reentered SDW gap. However the second SDW phase was not observed yet in the phase diagram. Also the hump-dip feature is observed in both electron and hole doped cases, which is hard to understand if there is only two SDW phases. Moreover this theory also finds difficulties in explaining the asymmetric line shape (one coherence peak is missing). 

(c) SC gap features. Based on the Eliashberg theory calculation,\cite{Schrieffer63} the SC gap function $\Delta(\omega)$ has some singularities around the Debey frequency $\omega\sim\omega_D$, which will be reflected in the STS spectrum $\mathrm{d}I/\mathrm{d}V\propto V/(V^2-\Delta(V)^2)^{1/2}$, represented as the little peaks/dips outside the SC coherence peak. This may provide an explanation for the extra peaks observed in the SC phase, but gives no understanding to the hump-dip feature in the normal phase.

(d) Inelastic electron tunneling (IET). In the STS experiment, if the electron tunnels into the sample with high enough energy, it may excite some gaped bosonic mode and opens a new tunneling channel, causing the differential conductance to jump up. Such a phenomenon is known as the IET,\cite{Hahn00} which may account for the observed hump-dip feature. In the context of iron-based superconductors, the relevant bosonic modes include magons, spin-resonance modes, and spinons. Our composite fermion explanation belongs to this picture, that the electron transfers its energy to the spinon to form a composite fermion that carries on the electric current. It is also possible for the electron to couple with the magon, but the resulting charge carrier would be a  magnetic polaron: the collective excitation of electron dressed  by a cloud of magnons. The polaron effect will enhance the electron mass (like the mass renormalization observed in ARPES), and cause an accumulation of the electron density of states around the Fermi level, which is opposite with the observe hump-dip feature. Another possibility is that the IET energy was transferred to the spin-resonance mode, which is fine for the extra peak in the SC phase, but hard to explain the hump-dip feature that persists into the normal phase.

Given the above difficulties in various theoretical scenarios, we conceived the composite fermion picture to understand the unusual experimental observations. The composite fermion may sounds exotic, but it is indeed a natural consequence that follows from the two-fluid description. Admittedly, we still do not have direct evidence for the existence of the deconfined spin-liquid state in the iron-based compounds. However as a candidate theory, we have shown that the two-fluid description was able to give account for various physics in the SDW, SC and normal phases systematically. We also proposed that the composite fermion mode can be tested by studying the magnetic field dependence of the hump-dip feature in the normal phase and the extra peak in the SC phase. More efforts along this direction may help to provide a more comprehensive understanding of the mechanism of iron-based superconductors.

\section{Conclusion}

In this work, we proposed the two-fluid description for iron-based superconductors.  In this scenario, the iron-based SC is understood neither as the weakly correlated BCS theory nor as the strongly correlated RVB theory, but a combination of both physics in some sense: the BCS pairing of itinerant electrons mediated by the fluctuations of the local moment spin-liquid background. The intermediate correlated electron system in the iron-based compounds contains two effective fluid components: the Femi-liquid of itinerant electrons and the spin-liquid of local moments. The separation of the two-fluid components seems to complicate the problem, but it in fact reduces the difficulty for theoretical handling, because both the Fermi-liquid and the spin-liquid physics have been well explored separately, thus the combined description could provide us a more definite starting point to understand intermediate correlated system.

The combination of the two-fluid components is not a simple superposition, as the Fermi-liquid and spin-liquid are coupled together via the renormalized Hund's rule interaction among the Fe $3d$ orbitals. The coupling leads to more interesting low-energy collective modes of both fluids, which include the bosonic magnon mode and the composite fermion mode. These modes are responsible for a series of low-energy phenomena in the iron-based compounds, including the SDW and SC ordering, the spin-resonance, and the hump-dip feature in the normal phase. To some extent, it is these emergent collective modes that constitute the new physics of intermediate correlated system that connect between the weakly or strongly correlated limits.

The two-fluid description assumes that the local moments in the iron-based superconductor are in a spin-liquid state with deconfined spinon excitations. For an isolated spin-liquid, the spinons are simply created and annihilated in pairs, driven by the strong quantum fluctuation. However coupling the spin-liquid to the Fermi-liquid allows the spinon to combine with the itinerant electron into the composite fermion, which promotes the spinon creation. In this sense, the fractionalization of the local moment is facilitated by its coupling to the itinerant electron, and the formation of composite fermion in turn justifies our initial assumption of the deconfined spin-liquid. This mechanism of the electron-facilitated fractionalization not only points out a prospective direction to search for spin-liquid states, but also provides a different way to study the spinons: the spinon is now labeled with charge by the attachment of the electron, and the resulting composite fermion mode can be easily probed by the electromagnetic response in various condensed matter experiments. If this scenario of the two-fluid description was correct, then the significance of the iron-based superconductors would not be limited to the question of high-temperature superconductivity, but could also be extended to the research of spin-liquid by providing new materials and measurements. 

\begin{acknowledgments}
We acknowledge the helpful discussions with Hong Yao, Yang Qi, Fan Yang, Tao Li, Fa Wang, Ya-Yu Wang, and Xiao-Dong Zhou. The work is supported by NBRPC (Nos. 2009CB929402, 2010CB923003).
\end{acknowledgments}

\onecolumngrid
\appendix
\section{Square Lattice PSG of $\mathbb{Z}_2$ Bosonic Spin-Liquid}\label{app: PSG}
In this appendix, we provide the PSG classification of bosonic $\mathbb{Z}_2$ spin-liquid on the square lattice, and then analyze its constraint on the mean-field ansatz.

We start from the symmetry group (SG) of the square lattice, which is generated by the following operations: the translation along $x$-direction $T_1: (x,y)\to(x+1,y)$, the translation along $y$-direction $T_2: (x,y)\to(x,y+1)$, the reflection about $x$-axis $\sigma_s: (x,y)\to(x,-y)$, the reflection about the diagonal $\sigma_d: (x,y)\to(y,x)$. Here $(x,y)$ label the lattice coordinates. For FeAs/FeSe layer in the iron-based compound, because of the alternating arrangement of the As/Se atoms outside the Fe-plane, the operations $T_1$, $T_2$ and $\sigma_s$ are actually implicitly followed by a mirror reflection about the Fe-plane. The symmetry group generators satisfies the following definition relations: $T_2 T_1 = T_1 T_2$, $\sigma_s T_1 = T_1 \sigma_s$, $\sigma_s T_2 = T_2^{-1} \sigma_s$, $\sigma_d T_1 = T_2 \sigma_d$, $\sigma_d T_2 = T_1 \sigma_d$, $\sigma_s^2 = \sigma_d^2 = (\sigma_s\sigma_d)^4 = 1$.

Under the action of $g\in$SG, the spinon $b_i$ is transformed as $b_i\to G_{g}(i)b_{g(i)}$, which contains a coordinate transform $i\to g(i)$ followed by a gauge transform $G_g(i)$ in the U(1) gauge group of the bosonic spinon. For $\mathbb{Z}_2$ spin-liquid, the invariant gauge group (IGG) is the $\mathbb{Z}_2$ group. To represent the definition relations, we introduce $\mathbb{Z}_2$ variables $p_i=0,1$ (for $i=1,\cdots,8$), then the algebraic PSG equations are
\begin{equation}
\begin{split}
G_{T_2}(x,y)G_{T_1}(x,y-1) &= (-)^{p_1} G_{T_1}(x,y) G_{T_2}(x-1,y),\\
G_{\sigma_s}(x,y)G_{T_1}(x,-y) &= (-)^{p_2} G_{T_1}(x,y) G_{\sigma_s}(x-1,y),\\
G_{\sigma_s}(x,y)G_{T_2}(x,-y) &= (-)^{p_3} G_{T_2}^{-1}(x,y+1)G_{\sigma_s}(x,y+1),\\
G_{\sigma_d}(x,y)G_{T_1}(y,x) &= (-)^{p_4} G_{T_2}(x,y) G_{\sigma_d}(x,y-1),\\
G_{\sigma_d}(x,y)G_{T_2}(y,x) &= (-)^{p_5} G_{T_1}(x,y)G_{\sigma_d}(x-1,y),\\
G_{\sigma_s}(x,y)G_{\sigma_s}(x,-y) &= (-)^{p_6},\\
G_{\sigma_d}(x,y)G_{\sigma_d}(y,x) &= (-)^{p_7},\\
G_{\sigma_s}(x,y)G_{\sigma_d}(x,-y)G_{\sigma_s}(-y,x)G_{\sigma_d}(-y,-x)&G_{\sigma_s}(-x,-y)G_{\sigma_d}(-x,y)G_{\sigma_s}(y,-x)G_{\sigma_d}(y,x) = (-)^{p_8}.
\end{split}
\end{equation}
One set of gauge inequivalent solutions of the above equations is given as follows
\begin{equation} \label{eq: PSG}
\begin{split}
G_{T_1}(x,y)&=1,\\
G_{T_2}(x,y)&=(-)^{p_{xy} x},\\
G_{\sigma_s}(x,y)&=(-)^{p_x x+p_y y}i^{p_s},\\
G_{\sigma_d}(x,y)&=(-)^{p_{xy} x y} i^{p_d}.
\end{split}
\end{equation}
The solutions are classified by 5 indices $p_{xy},p_x,p_y,p_s,p_d=0,1$, which leads to totally $2^5=32$ classes of algebraic PSG's.

Now we wish to find the mean-field ansatz for the symmetric spin-liquid close to the $(\pi,0)$ ordering. The ordering favors the nearest neighboring (nn) hopping $\chi_1$ and the next nearest neighboring (nnn) pairing $\eta_2$. However in the presence of both $\chi_1$ and $\eta_2$, the nn pairing $\eta_1$ would also be induced. So we seek for a PSG that allows simultaneous presence of $\chi_1$, $\eta_1$ and $\eta_2$.

First consider the presence of nnn pairing $\eta_2$ between the sites (0,0) and (1,1). The bound direction gets reversed under $\sigma_s T_1 T_2^{-1}\sigma_d\sigma_s$, so we must have $\eta_2\to-\eta_2$ under the PSG action. Given the solution in \eqnref{eq: PSG}, we find $\eta_2\to(-)^{p_d}\eta_2$ under the PSG action of $\sigma_s T_1 T_2^{-1}\sigma_d\sigma_s$, thus $p_d=1$. Moreover, the operation $\sigma_d$ leaves the diagonal bound unchanged, while under the corresponding PSG action, $\eta_2\to(-)^{p_d+p_{xy}}\eta_2$ according to \eqnref{eq: PSG}, so we must have $(-)^{p_d+p_{xy}}=1$, which leads to $p_{xy}=1$. Now consider the presence of nn hopping $\chi_1$ between the sites (0,0) and (0,1). The bound direction gets reversed under $T_2\sigma_s$, while the corresponding PSG action takes $\chi_1\to (-)^{p_y}\chi_1$, so we must have $\chi_1^*=(-)^{p_y}\chi_1$. On the other hand, the operation $\sigma_d\sigma_s\sigma_d$ leaves the vertical bound unchanged, while under the corresponding PSG action, $\chi_1\to(-)^{p_x}\chi_1$, thus we must have $(-)^{p_x}=1$, from which $p_x=0$. At last, we consider the presence of nn pairing $\eta_1$ between the sites (0,0) and (1,0). Under the PSG action of $T_2\sigma_s$: $\eta_1\to(-)^{p_s-p_y}\eta_1$, while the bound direction is reversed, so we must have $(-)^{p_s-p_y}=-1$. Also, under the PSG action of $\sigma_d\sigma_s\sigma_d$, $\eta_1\to(-)^{p_s+p_x}\eta_1$, while the bound is untouched, so we must have $(-)^{p_s+p_x}=1$. Given $p_x=0$, we have $p_s=0$. Then according to $(-)^{p_s-p_y}=-1$, we conclude $p_y=1$. So the equation $\chi_1^*=(-)^{p_y}\chi_1=-\chi_1$ requires nn hoping to be pure imaginary. In conclusion, the only PSG that allows simultaneous presence of nn hoping and pairing, and nnn pairing is as follows
\begin{equation}\label{eq: PSG0}
\begin{split}
G_{T_1}(x,y)&=1,\\
G_{T_2}(x,y)&=(-)^x,\\
G_{\sigma_s}(x,y)&=(-)^{y},\\
G_{\sigma_d}(x,y)&=i(-)^{x y}.
\end{split}
\end{equation}
This PSG also requires the nn hoping to be pure imaginary, which may be re-parameterized by $i\chi_1$ (such that $\chi_1\in \mathbb{R}$). We may also set $\eta_2\in\mathbb{R}$. Then use the PSG given in \eqnref{eq: PSG0} to send the mean-field parameters to all bounds in the lattice, the eventual result is given in \eqnref{eq: MF ansatz}.


\begin{thebibliography}{99}
\bibitem{Hosono08} Y. Kamihara, T. Watanabe, M. Hirano, H. Hosono, J. Am. Chem. Soc. {\bf130}, 3296 (2008).
\bibitem{BCS} J. Bardeen, L. N. Cooper, J. R. Schieffer, Phys. Rev. {\bf 108}, 1175 (1957); Phys. Rev. {\bf106}, 162 (1957).
\bibitem{RVB} P. W. Anderson, Science {\bf235}, 1196 (1987).
\bibitem{Mazin09} I. I. Mazin, J. Schmalian, Physica C {\bf469}, 614 (2009). 
\bibitem{Zhao09} J. Zhao, D. T. Adroja, D.-X. Yao, R. Bewley, S. Li, X. F. Wang, G. Wu, X. H. Chen, J. Hu, and P. Dai, Nat. Phys. {\bf5}, 555 - 560 (2009).
\bibitem{Hansmann10}P. Hansmann, R. Arita, A. Toschi, S. Sakai, G. Sangiovanni, and K. Held, Phys. Rev. Lett. {\bf104}, 197002 (2010).
\bibitem{Gretarsson11}H. Gretarsson, A. Lupascu, Jungho Kim, D. Casa, T. Gog, W. Wu, S. R. Julian, Z. J. Xu, J. S. Wen, G. D. Gu, R. H. Yuan, Z. G. Chen, N.-L. Wang, S. Khim, K. H. Kim, M. Ishikado, I. Jarrige, S. Shamoto, J.-H. Chu, I. R. Fisher, and Y.-J. Kim, Phys. Rev. B {\bf84}, 100509(R) (2011).
\bibitem{Vilmercati12}P. Vilmercati, A. Fedorov, F. Bondino, F. Offi, G. Panaccione, P. Lacovig, L. Simonelli, M. A. McGuire, A. S. M. Sefat, D. Mandrus, B. C. Sales, T. Egami, W. Ku, and N. Mannella, Phys. Rev. B {\bf85}, 220503(R) (2012).
\bibitem{Gorkov13}L. P. Gor'kov and G. B. Teitel'baum, Phys. Rev. B {\bf87}, 024504 (2013).
\bibitem{Medici09} L. de'Medici, S.~R. Hassan, M. Capone, and X. Dai, Phys. Rev. Letts. {\bf 102}, 126401  (2009).
\bibitem{Lv10} W. Lv, F. Kr\"uger, P. Phillips, Phys. Rev. B {\bf 82}, 045125 (2010). 
\bibitem{Medici11} L. de'Medici, Phys. Rev. B {\bf 83},  205112  (2011).
\bibitem{Yu11}R. Yu and Q. Si, Phys. Rev. B {\bf 84},  235115  (2011).
\bibitem{Zhang:2012bh}Y.-Z. Zhang, H. Lee, H.-Q. Lin, C.-Q. Wu, H. O. Jeschke, R. Valenti, Phys. Rev. B {\bf 85},  035123  (2012).
\bibitem{Quan:2012dq} Y.-M. Quan, L.-J. Zou, D.-Y. Liu, and H.-Q. Lin, Euro. Phys. J. B {\bf 85},  1 (2012).
\bibitem{Yu:2012ve} R. Yu and Q. Si, arXiv:1208.5547 (2012).
\bibitem{Moon10} S. J. Moon, J. H. Shin, D. Parker, W. S. Choi, I. I. Mazin, Y. S. Lee, J. Y. Kim, N. H. Sung, B. K. Cho, S. H. Khim, J. S. Kim, K. H. Kim, and T. W. Noh, Phys. Rev. B {\bf81}, 205114 (2010).
\bibitem{NLWang12} N.-L. Wang, W.-Z. Hu, Z.-G. Chen, R.-H. Yuan, G. Li, G.-F. Chen, and T. Xiang, J. Phys. Condens. Mat. {\bf24}, C4202 (2012).
\bibitem{Weng09} Z.-Y. Weng, Physica E {\bf41}, 1281 (2009).
\bibitem{KouLiWeng09} S.-P. Kou, T. Li, and Z.-Y. Weng, Europhys. Letts. {\bf88}, 17010 (2009).
\bibitem{Yin10} W.-G. Yin, C.-C. Lee, and W. Ku, Phys. Rev. Letts. {\bf105},107004 (2010). 
\bibitem{YouPRB11}Y.-Z. You, F. Yang, S.-P. Kou, Z.-Y. Weng, Phys. Rev. B {\bf84}, 054527 (2011).
\bibitem{YYWang12} X. Zhou, P. Cai, A. Wang, W. Ruan, C. Ye, X. Chen, Y. You, Z.-Y. Weng, Y. Wang, Phys. Rev. Lett. {\bf 109}, 037002 (2012). 
\bibitem{Wang2013} Z. Wang, H. Yang, D. Fang, B. Shen, Q.-H. Wang, L. Shan, C. Zhang, P. Dai, H.-H. Wen, Nat. Phys. {\bf9}, 42-48 (2013).
\bibitem{Wen91} X.-G. Wen, Phys. Rev. B {\bf 44}, 2664 (1991).
\bibitem{ReadSachdev91} N. Read and S. Sachdev, Phys. Rev. Lett. {\bf 66}, 1773 (1991); S. Sachdev, N. Read, Int. J.  Mod. Phys. B {\bf 5}, 219 (1991).
\bibitem{Sachdev94} A. V. Chubukov, T. Senthil, S. Sachdev, Phys. Rev. Letts. {\bf72}, 2089 (1994).
\bibitem{Senthil04} T. Senthil, L. Balents, S. Sachdev, A. Vishwanath, M. P. A. Fisher, Phys. Rev. B {\bf 70}, 144407 (2004).
\bibitem{Arovas88} D. P. Arovas, A. Auerbach, Phys. Rev. B {\bf38}, 316 (1988). 
\bibitem{Davis10}T.-M. Chuang, M. P. Allan, J. Lee, Y. Xie, Ni Ni, S. L. Budko, G. S. Boebinger, P. C. Canfield, and J. C. Davis, Science {\bf327}, 181 (2010).
\bibitem{Fisher10}J.-H. Chu, J. G. Analytis, K. de Greve, P. L. McMahon, Z. Islam, Y. Yamamoto, and I. R. Fisher, Science {\bf329}, 824 (2010).
\bibitem{Shen11}M. Yi, D. Lu, J.-H. Chu, J. G. Analytis, A. P. Sorini, A. F. Kemper, B. Moritz, S.-K. Mo, R. G. Moore, M. Hashimoto, W.-S. Lee, Z. Hussain, T. P. Devereaux, I. R. Fisher, and Z.-X. Shen, Proc. Natl. Acad. Sci. USA {\bf108}, 6878 (2011).
\bibitem{Xue11}C.-L. Song, Y.-L. Wang, P. Cheng, Y.-P. Jiang, W. Li, T. Zhang, Z. Li, K. He, L. Wang, J.-F. Jia, H.-H. Hung, C. Wu, X. Ma, X. Chen, and Q.-K. Xue, Science {\bf332}, 1410 (2011).
\bibitem{Matsuda12}S. Kasahara, H. J. Shi, K. Hashimoto, S. Tonegawa, Y. Mizukami, T. Shibauchi, K. Sugimoto, T. Fukuda, T. Terashima, Andriy H. Nevidomskyy, and Y. Matsuda, Nature (London) {\bf486}, 382 (2012).
\bibitem{Fisher12}J.-H. Chu, H.-H. Kuo, J. G. Analytis, and I. R. Fisher, Science {\bf337}, 710 (2012).
\bibitem{FYang2013}F. Yang, F. Wang, and D.-H. Lee, Phys. Rev. B 88, 100504(R) (2013).
\bibitem{Si08} Q. Si, E. Abrahams, Phys. Rev. Lett. {\bf101}, 076401 (2008). 
\bibitem{MaLuXiang08} F. Ma, Z.-Y. Lu, T. Xiang, Phys. Rev. B {\bf78}, 224517 (2008). 
\bibitem{Sigh10} R. Applegate, J. Oimaa, R. R. P. Sigh, Phys. Rev. B {\bf81}, 024505 (2010). 
\bibitem{Si11} R. Yu, P. Goswami, Q. Si, Phys. Rev. B {\bf 84}, 094451 (2011). 
\bibitem{Cruz08} C. de la Cruz, Q. Huang, J. W. Lynn, J. Li, W. Ratcliff II, J. L. Zarestky, H. A. Mook, G. F. Chen, J. L. Luo, N. L. Wang, P. Dai, Nature, {\bf453}, 899 (2008). 
\bibitem{Huang08} Q. Huang, Y. Qui, W. Bao, M. A. Green, J. W. Lynn, Y. C. Gasparovic, T. Wu, G. Wu, X. H. Chen, Phys. Rev. Lett. {\bf101}, 257003 (2008).
\bibitem{Li09} S. Li, C. de la Cruz, Q. Huang, G. F. Chen, T.-L. Xia, J. L. Luo, N. L. Wang, P. Dai, Phys. Rev. B {\bf80}, 020504 (2009).
\bibitem{Richter08} R. Darradi, O. Derzhko, R. Zinke, J. Schulenburg, S. E. Kr\"uger, J. Richter, Phys. Rev. B {\bf 78}, 214415 (2008). 
\bibitem{Jiang09} H. C. Jiang, F. Kr\"uger, J. E.  Moore, D. N. Sheng, J. Zaanen, Z. Y. Weng, Phys. Rev. B {\bf79}, 174409 (2009). 
\bibitem{Richter10} J. Richter, J. Schulenburg, Eur. Phys. J. B {\bf73}, 117-124 (2010).  
\bibitem{Reuther10} J. Reuther, W. Peter, Phys. Rev. B {\bf 81}, 144410 (2010). 
\bibitem{Jiang11} H.-C. Jiang, H. Yao, L. Balents, Phys. Rev. B {\bf 86}, 024424 (2012). 
\bibitem{LWang11} L. Wang, Z.-C. Gu, F. Verstraete, X.-G. Wen, arXiv:1112.3331. 
\bibitem{TLi12} T. Li, F. Becca, W. Hu, S. Sorella, Phys. Rev. B {\bf 86}, 075111 (2012). 
\bibitem{YangYao12} F. Yang, H. Yao, Phys. Rev. Lett. {\bf 109}, 147209 (2012). 
\bibitem{Yin12} W.-G. Yin, C.-C. Lee, and W. Ku, Supercond. Sci. Technol. {\bf25}, 084007  (2012).
\bibitem{Yao10} H. Yao, L. Fu, X.-L. Qi, arXiv:1012.4470
\bibitem{Doniach79} S. Doniach, Physica B {\bf91}, 231 (1979).
\bibitem{Lacroix79} C. Lacroix and M. Cyrot,  Phys. Rev. B {\bf20}, 1969 (1979).
\bibitem{Nakatsuji04} S. Nakatsuji, D. Pines, and Z. Fisk, Phys. Rev. Letts. {\bf92}, 016401 (2004).
\bibitem{Yang08hc} Y.-F. Yang and D. Pines, Phys. Rev. Letts. {\bf100}, 096404 (2008).
\bibitem{Yang08bs} Y.-F. Yang, Z. Fisk, H.-O. Lee, J. D. Thompson, and D. Pines, Nature {\bf454}, 611 (2008). 
\bibitem{Anderson70}P. W. Anderson, J. Phys. C: Solid State Phys. {\bf3}, 2436 (1970).
\bibitem{Mazin10} I. I. Mazin, Natrue {\bf464}, 183 (2010).
\bibitem{Hirschfeld11} P. J. Hirschfeld, M. M. Kroshunov, I. I. Mazin, Rep. Prog. Phys. {\bf74}, 124508 (2011)
\bibitem{Chubukov08} A. V. Chubukov, D. V. Efremov, I. Eremin, Phys. Rev. B {\bf 78}, 134512 (2008). 
\bibitem{Hu12} J. Hu, N. Hao, Phys. Rev. X {\bf2}, 021009 (2012); J. Hu, arXiv:1208.6201. 
\bibitem{MaLu08} F. Ma, Z.-Y. Lu, Phys. Rev. B {\bf 78}, 033111 (2008). 
\bibitem{Singh08} D. J. Singh, M.-H. Du, Phys, Rev. Lett. {\bf100}, 237003 (2008). 
\bibitem{Singh09} L. Zhang, D. J. Singh, Phys. Rev. B {\bf79}, 174530 (2009).
\bibitem{FeSe1} Y. Zhang, L. X. Yang, M. Xu, Z. R. Ye, F. Chen, C. He, H. C. Xu, J. Jiang, B. P. Xie, J. J. Ying, X. F. Wang, X. H. Chen, J. P. Hu, M. Matsunami, S. Kimura, D. L. Feng, Nature Mater. {\bf10}, 273 (2011).
\bibitem{FeSe2} D. Mou, S. Liu, X. Jia, J. He, Y. Peng, L. Zhao, L. Yu, G. Liu, S. He, X. Dong, J. Zhang, H. Wang, C. Dong, M. Fang, X. Wang, Q. Peng, Z. Wang, S. Zhang, F. Yang, Z. Xu, C. Chen, and X. J. Zhou, Phys. Rev. Lett. {\bf106}, 107001 (2011).
\bibitem{FeSe3}X.-P. Wang, T. Qian, P. Richard, P. Zhang, J. Dong, H.-D. Wang, C.-H. Dong, M.-H. Fang and H. Ding, Europhys. Lett. {\bf93}, 57001 (2011).
\bibitem{Goldman08}A. I. Goldman, D. N. Argyriou, B. Ouladdiaf, T. Chatterji, A. Kreyssig, S. Nandi, N. Ni, S. L. Bud'ko, P. C. Canfield, R. J. McQueeney, Phys. Rev. B {\bf78}, 100506(R) (2008).
\bibitem{Kaneko08}K. Kaneko, A. Hoser, N. Caroca-Canales, A. Jesche, C. Krellner, O. Stockert, C. Geibel, Phys. Rev. B {\bf78}, 212502 (2008).
\bibitem{Matan09}K. Matan, R. Morinaga, K. Iida, T. J. Sato, Phys. Rev. B {\bf79}, 054526 (2009).
\bibitem{Si12} R. Yu, Z. Wang, P. Goswami, A. H. Nevidomskyy, Q. Si, E. Abrahams, Phys. Rev. B {\bf86}, 085148 (2012). 
\bibitem{Sachdev91} S. Sachdev, Phys. Rev. B {\bf 45} 12377 (1991). 
\bibitem{WangVishwanath} F. Wang and A. Vishwanath, Phys Rev. B. {\bf 74}, 174423 (2006)
\bibitem{Wang10} F. Wang, Phys. Rev. B {\bf 82}, 024419 (2010).
\bibitem{Wen02} X.-G. Wen, Phys. Rev. {\bf B} 65, 165113 (2002).
\bibitem{Zhai09} H. Zhai, F. Wang, D.-H. Lee, Phys. Rev. B {\bf80}, 064517 (2009). 
\bibitem{Dolgov09}O. V. Dolgov, I. I. Mazin, D. Parker, A. A. Golubov, Phys. Rev. B {\bf79}, 060502 (2009). 
\bibitem{Manhan} G. D. Mahan, Many-Particle Physics 3rd Ed., Chap. 10.1, p. 628-644, Kluwer Academic/Plenum Publishers, New York, 2000. 
\bibitem{Scalapino08a} T. A. Maier and D. J. Scalapino, Phys. Rev. B {\bf 78}, 020514(R) (2008). 
\bibitem{Scalapino08b} M. Daghofer, A. Moreo, J. A. Riera, E. Arrigoni, D. J. Scalapino, and E. Dagotto, Phys. Rev. Lett. {\bf101}, 237004 (2008). 
\bibitem{Korshunov08} M. M. Korshunov and I. Eremin, Phys. Rev. B {\bf 78}, 140509(R) (2008). 
\bibitem{Bernevig09} K. Seo, C. Fang, B. A. Bernevig, and J. Hu, Phys. Rev. B {\bf79}, 235207 (2009). 
\bibitem{Christianson08} A. D. Christianson, E. A. Goremychkin, R. Osborn, S. Rosenkranz, M. D. Lumsden, C. D. Malliakas, I. S. Todorov, H. Claus, D. Y. Chung, M. G. Kanatzidis, R. I. Bewley and T. Guidi, Nature {\bf456}, 930-932 (2008). 
\bibitem{Lumsden09}M. D. Lumsden, A. D. Christianson, D. Parshall, M. B. Stone, S. E. Nagler, G. J. MacDougall, H. A. Mook, K. Lokshin, T. Egami, D. L. Abernathy, E. A. Goremychkin, R. Osborn, M. A. McGuire, A. S. Sefat, R. Jin, B. C. Sales, and D. Mandrus, Phys. Rev. Lett. {\bf102}, 107005 (2009). 
\bibitem{Qui09}Y. Qiu, W. Bao, Y. Zhao, C. Broholm, V. Stanev, Z. Tesanovic, Y. C. Gasparovic, S. Chang, J. Hu, B. Qian, M. Fang, and Z. Mao, Phys. Rev. Lett. {\bf103}, 067008 (2009).
\bibitem{Taylor11}A. E. Taylor, M. J. Pitcher, R. A. Ewings, T. G. Perring, S. J. Clarke, and A. T. Boothroyd, Phys. Rev. B {\bf83}, 220514(R) (2011). 
\bibitem{Christianson09}A. D. Christianson, M. D. Lumsden, S. E. Nagler, G. J. MacDougall, M. A. McGuire, A. S. Sefat, R. Jin, B. C. Sales, and D. Mandrus, Phys. Rev. Lett. {\bf103}, 087002 (2009).
\bibitem{Chubukov10}A. B. Vorontsov, M. G. Vavilov, and A. V. Chubukov, Phys. Rev. B {\bf81}, 174538 (2010).
\bibitem{YKW10}F. Yang, S.-P. Kou, and Z.-Y. Weng, Phys. Rev. B {\bf81}, 245130 (2010).
\bibitem{PDai12} M. Liu, L. W. Harriger, H. Luo, M. Wang, R. A. Ewings, T. Guidi, H. Park, K. Haule, G. Kotliar, S. M. Hayden, and P. Dai, Nature Physics {\bf8}, 376-381 (2012).
\bibitem{Liu10} C. Liu, Y. Lee, A. D. Palczewski, J.-Q. Yan, T. Kondo, B. N. Harmon, R. W. McCallum, T. A. Lograsso, and A. Kaminski, Phys. Rev. B {\bf82}, 075135 (2010).
\bibitem{Liu11} Z.-H. Liu, P. Richard, K. Nakayama, G.-F. Chen, S. Dong, J.-B. He, D.-M. Wang, T.-L. Xia, K. Umezawa, T. Kawahara, S. Souma, T. Sato, T. Takahashi, T. Qian, Y. Huang, N. Xu, Y. Shi, H. Ding, and S.-C. Wang, Phys. Rev. B {\bf84}, 064519 (2011).
\bibitem{Richard11} P. Richard, T. Sato, K. Nakayama, T. Takahashi, and H. Ding, Rep. Prog. Phys. {\bf74}, 124512 (2011).
\bibitem{Zhang12} Y. Zhang, C. He, Z. R. Ye, J. Jiang, F. Chen, M. Xu, Q. Q. Ge, B. P. Xie, J. Wei, M. Aeschlimann, X. Y. Cui, M. Shi, J. P. Hu, and D. L. Feng, Phys. Rev. B {\bf85}, 085121 (2012).
\bibitem{Jiang12} H.-M. Jiang and Z.-J. Yao and F.-C. Zhang, Euro. Phys. Letts. {\bf100}, 47004 (2012).
\bibitem{Schrieffer63} J. R. Schrieffer, D. J. Scalapino, J. W. Wilkins, Phys. Rev. Letts. {\bf10}, 336 (1963).
\bibitem{Hahn00} J. R. Hahn, H. J. Lee, and W. Ho, Phys. Rev. Letts. {\bf85}, 1914 (2000).
\end{thebibliography}
\end{document}